\documentclass[aps,prd,onecolumn,showpacs,preprintnumbers]{revtex4}

\usepackage{graphicx}
\usepackage{amssymb}

\begin{document}

 
%\preprint{IUCAA-??}
%\preprint{IISc-CTS-1/03}
\title{Role of Horizons in Semiclassical Gravity:
       Entropy and the Area Spectrum}

  \author{T. Padmanabhan}\email{nabhan@iucaa.ernet.in}
  \affiliation{Inter-University Center for Astronomy and Astrophysics,
  Post Bag 4, Ganeshkhind, Pune-411007}
  \author{Apoorva Patel}\email{adpatel@cts.iisc.ernet.in}
  \affiliation{CTS and SERC, Indian Institute of Science, Bangalore-560012}


\date{\today}

\begin{abstract}

In any space-time, it is possible to have a family of observers who have
access to only part of the space-time manifold, because of the existence
of a horizon.
We demand that
\emph{physical theories in a given coordinate system must be formulated
entirely in terms of variables that an observer using that coordinate system
can access}.
In the coordinate frame in which these observers are at rest, the horizon
manifests itself as a (coordinate) singularity in the metric tensor.
Regularization of this singularity removes the inaccessible region,
and leads to the following consequences:
(a) The non-trivial topological structure for the effective manifold
allows one to obtain the standard results of quantum field theory
in curved space-time.
(b) In case of gravity, this principle requires that the effect of the unobserved degrees of
freedom should reduce to a boundary contribution $A_{\rm boundary}$ to the gravitational action.
When the boundary is a horizon, $A_{\rm boundary}$ reduces 
to a single, well-defined term proportional to the area of the horizon.
Using the form of this boundary term, it is possible to obtain the
full gravitational action in the semiclassical limit.
(c) This boundary term must have a quantized spectrum with uniform
spacing, $\Delta A_{boundary}=2\pi\hbar$, in the semiclassical limit.
This, in turn, yields the following results for semiclassical gravity:
(i) The area of any one-way membrane is quantized.
(ii) The information hidden by a one-way membrane amounts to an entropy,
which is always one-fourth of the area of the membrane in the leading order.
(iii) In static space-times, the action for gravity can be given a purely
thermodynamic interpretation and the Einstein equations have a formal
similarity to laws of thermodynamics.    
\end{abstract}

%\keywords{}

\maketitle


\section{Introduction}
\subsection{Motivation}

Consider a congruence of time-like curves in a region of space-time,
which could represent a family of observers in that space-time.
Any such family of observers can study
physics in their own frame of reference, and arrive at conclusions
which can be  translated between different families of observers.
In attempting to implement this criterion, one needs to consider 
whether the observers have access to the same region of space-time or not.
We can construct the union of the causal past (essentially the
union of all the backward light cones) of all events in this congruence.
If this union has a non-trivial boundary $\mathcal{H}$,
then $\mathcal{H}$ will act as a horizon limiting the region
from which these observers can receive information.
It is trivial to construct time-like congruences with such non-trivial
causal boundaries, in any space-time.
(Examples: the uniformly accelerated observers in flat
space-time;  observers with fixed Schwarzschild coordinates in Kruskal
space-time.)
We need to examine the question of how the observers following trajectories
on these congruences will perceive different physical phenomena.

The family of observers can be provided with a natural coordinate system
$(t,{\bf x})$, which---quite minimally---corresponds to a system in which
each trajectory of the congruence is mapped to a ${\bf x}$=constant line.
The observers are then at rest in this coordinate system.
This does not uniquely fix the coordinate system (in particular,
it does not fix the time coordinate), but it is sufficient for our purpose.
(Examples: Rindler frame with coordinates $(t,l,y,z)$ on the flat
space-time manifold with metric $(-g_{00}=1/g_{11}=al,g_{yy}=g_{zz}=1)$;
 the Schwarzschild frame $(t,r,\theta,\phi)$ on the Kruskal manifold
with metric $(-g_{00}=1/g_{11}=1-(2M/r))$.)
By considering trajectories of the congruence which are close to $\mathcal{H}$,
it is easy to convince oneself that the metric tensor in this coordinate
system will behave badly on $\mathcal{H}$.
In fact, it is this ``bad'' behaviour of the metric tensor components,
at events which do not have curvature singularities, that alerts one to
the existence of horizons.
(Examples: the $l=0$ surface in case of the Rindler frame;
the $r=2M$ surface in case of the Schwarzschild frame.)
In all these cases, the horizon is defined with respect to a pre-specified
congruence of time-like curves, but in the same space-time there also exist
other time-like congruences which are not limited by a horizon.
(Examples: inertial observers in flat space-time;  observers falling
in to a black-hole.)

The existence of a class of observers with limited access to space-time
regions, because of the existence of horizons, is a generic feature.
It has nothing to do with the dynamics of general relativity or gravity;
such examples exist even in flat space-time.
But when the space-time is flat, one can introduce an additional ``rule''
that only inertial coordinates be used to describe physics.
While this appears to be artificial and ad hoc, it is logically tenable.
But gravitational interaction, which makes space-time curved, removes this
option and forces us to consider  curvilinear coordinate systems.
Since we cannot use globally inertial coordinates
to describe physics in general (because of the existence of gravity),
it makes sense to study the effects of curvilinear coordinates
even in flat space-time.
One can think of issues arising from curvilinear coordinates in a general
setting, without asking whether one is in flat or curved space-time.

The key point is that while we would like all families of observers
to describe their observations using the same theory, the regions of
space-time accessible to different families of observers can be different.
These two features impose conflicting demands on a theory,
and a theory that can reconcile them has to be highly restrictive.
The \emph{classical} general relativity falls in this category,
because of its high symmetry and the specific structure of its action
functional.
In any attempt to quantize gravity, new structure is added to general
relativity, and consequences of the conflicting demands require reanalysis.

Difficulties arise already at the level of \emph{quantum} physics in
curvilinear coordinates, and these difficulties can be traced to the fact
that quantum field theoretical amplitude $G({\mathcal A},{\mathcal B})$ for
propagation of a particle between two events ${\mathcal A}$ and ${\mathcal B}$
is: (i) non-zero for events separated by a space-like interval in virtual
processes, and (ii) requires analytic continuation to complex values of the
time coordinate (either through $i\epsilon$-prescription or by Euclidean
extension) to restore causality for real (i.e. on-shell) particles.
This is a priori surprising because causality is related to the existence
of light cones and there is no notion of light cones in the Euclidean sector.
It works because special relativity has observer-independent light cone
structure in the inertial frames, and hence the singularities of the
wave operator can be identified and regulated using analytic continuation. 
When we move on to general relativity, light cone structure becomes
observer-dependent; a family of observers with a horizon will have to
interpret the amplitude for propagation of a particle across the horizon,
even though they have no access to the region beyond the horizon.
Furthermore, the analytic continuation $t\to t_E=it$ does not commute
with the general coordinate transformation $t\to f(t,{\bf x})$,
and different choices for analytic continuation can lead to different,
unitarily inequivalent, quantum field theories.
(A simple example is the conflict between analytic continuation in the
inertial time coordinate and that in the Rindler time coordinate.)
When the coordinate system has a horizon, the effect of the horizon in the Euclidean sector is to introduce
certain singularities which needs to be regularized afresh. It is, therefore,
 imperative that we try to understand quantum phenomena in curvilinear
coordinates from a broader perspective. 

\subsection{The principle of effective theory}

The aspect that different observers may have access to different regions of
space-time and hence differing amount of information, leads us to postulate
that \emph{physical theories in a given coordinate system must be formulated
entirely in terms of the variables that an observer using that coordinate
system can access}.
We call this postulate the ``principle of effective theory'', and it has
interesting consequences for both quantum field theory in curved space-time
and for the dynamics of gravity, which we explore in this paper.

This postulate is a new addition to the traditional principles of general
relativity, although it is a familiar principle in high energy physics.
In the context of field theories, a postulate of the above kind ``protects"
the low energy theories from unknown complications of the high energy sector.
For example, one can use QED to predict results at, say $10$ GeV,
without worrying about the structure of the theory at $10^{19}$ GeV,
as long as one uses coupling constants and variables
defined around $10$ GeV and determined observationally.
In this case, one invokes the effective field theory approach in momentum
space, and the effect of high-energy modes is essentially absorbed in the
definitions of field variables and coupling constants appearing in the low
energy Lagrangian.
The powerful formalism of renormalization group describes how a theory
changes as its domain of applicability is changed, and the symmetries of the
theory often provide a good guide to the types of changes that may occur.
We have learned from experience that in effective field theories
everything outside the domain accessible to the observer is reduced to:
(a) change of variables, (b) change of couplings,
(c) higher derivative interaction terms and (d) boundary terms.
(The number of interaction terms in the effective field theory may become
infinite, but they can be organized in increasing powers of derivatives).
Our postulate invokes the same reasoning in coordinate space (which is
commonplace in condensed matter physics and lattice field theories),
and demands---for example---that the observed physics outside a black hole
horizon must not depend on the unobservable processes beyond the horizon.
The effective field theory of gravity must be obtained by summing over
different configurations---and possibly different topologies---beyond the
horizon.
In such a theory, among the possible changes listed above,
(a) is unimportant because the geometrical description of gravity
identifies the metric tensor as the natural fundamental variable;
(b) is unimportant because the couplings are fixed using the effective theory;
and (c) is unimportant at the lowest order of effective theory.
Hence we concentrate on the boundary terms in the theory, i.e.
we expect the action functional describing gravity to contain certain
boundary terms which are capable of encoding the information equivalent
to that present beyond the horizon. 

While we introduce the principle of effective theory as a postulate,
we stress that it is a fairly natural demand.
To see this, let us recall that in the standard description of flat
space-time physics, one often divides the space-time by a space-like
surface $t=t_0$=constant.
With appropriate information on this surface, one can predict the
evolution for $t>t_0$, \emph{without knowing the details at} $t<t_0$.
In case of curved space-times with horizons, similar considerations apply.
For example, if the space-time contains a Schwarzschild black hole,
then the light cone structure guarantees that the processes inside the
black hole horizon cannot affect the outside events \emph{classically},
and our principle of effective theory is trivially realized.
But the situation in \emph{quantum theory} is more complicated because
quantum fields can have non-trivial correlations across the horizon
and---in general---can lead to processes which are classically forbidden.
The principle of effective theory requires that all such quantum effects
be encoded by a surface term. We shall see that this is indeed possible. 
 
\subsection{Outline of the paper}

As far as quantum field theory in a given manifold is concerned
(see Sections II,III and Appendix A), the principle of effective theory
requires that the theory should be formulated in an {\it effective manifold},
where the region inaccessible to the family of observers is removed.
Moreover, since the formulation of quantum field theory needs analytic
continuation to complex time, the effective manifold should have an
extension in the Euclidean sector.
In a wide class of space-times with horizon, the metric near the horizon
can be approximated as the Rindler metric.
(This is trivially true for many static horizons; the procedure works
even for a broader class of metrics which are conformally static.)
In the Euclidean sector, the entire region which is not accessible to the
family of observers reduces to a point.
The effective Euclidean manifold then has one point removed
from the whole manifold, and acquires non-trivial topology.
Standard results like the thermal behaviour of horizons etc.
arise from this feature at one go.
It was known \cite{chrisduff,vandam}
that introducing non-trivial topology by removing a point from the
Euclidean manifold leads to standard results, but it has never been clear
{\it why} this point should be removed \cite{tpinstanton}.
Indeed, it is important to formulate a consistent regularization of the
space-time near a horizon, to understand {\it how} this point has to be removed.
In the approach presented here, this procedure fits in to a larger picture,
and---more importantly---it sets the stage for the study of gravity. 

In case of gravity (which is our main concern), the principle of effective
theory not only determines the form of the gravitational action functional,
but also leads to specific results for the semiclassical limit of quantum
gravity. This is explored in Sections IV to VI.
The crucial new feature is that, once we remove the region inaccessible
to a family of observers, the information that is eliminated has to be
encoded in a boundary term of the theory.
One can fix the nature of the boundary term and then the full form
of the gravitational action, using some general considerations.
In turns out that
\emph{the boundary term is proportional to the area of the horizon}.
Semiclassical quantization of the theory then leads to certain constraints
on the action functional if the principle of effective theory is to be valid.
We show that, among other things,
this leads to the quantization of horizon areas.

The paper is organized as follows: Section II provides the background
describing the role of horizons and gauges in which the metric is singular.
We next discuss quantum field theory in singular gauges in Section III.
Since much of this is well known, in the body of the paper we concentrate
on conceptual issues and the role played by our paradigm,
and briefly describe the results in Appendix A.
Sections IV to VI discuss our approach in the context of semiclassical
gravity and derive new results.
Some of the technical details are given in Appendix B and C.
We use the space-time signature $(-,+,+,+)$ and other sign conventions of
ref.\cite{mtw}. Whenever not explicitly mentioned, our units are $G=\hbar=c=1$.
The Latin indices vary over 0-3, while the Greek indices cover 1-3.

\section{The Role of Horizons in physical theories}
\subsection{Horizon for a family of observers}

Consider a time-like trajectory $X^a(t)$ in a space-time,
parametrized by the proper time $t$ of the clock moving along that trajectory.
We can construct the past light cone $\mathcal{C}(t)$
for each event $\mathcal{P}[X^a(t)]$ on this trajectory.
The union $U$ of all these past light cones,
$\{\mathcal{C}(t),-\infty\leq t \leq \infty\}$,
determines whether an observer on the trajectory $X^a(t)$ can receive
information from all events in the space-time or not \cite{comment0}.
If $U$ has a non-trivial boundary, there will be regions of space-time
from which this observer cannot receive signals.
(We shall always use the term ``observer'' as synonymous to a time-like
trajectory in the space-time, without any other additional connotations.) 
In fact, one can extend this notion to a congruence of time-like curves
that fill a region of space-time.
Given a congruence of time-like curves (``family of observers"),
the boundary of the union of their causal pasts will define a (past)
{\it horizon} for this set of observers.
This horizon is dependent on the family of observers,
but is coordinate independent.
We stress that a horizon $\mathcal{H}$ is defined with respect to a
\emph{specific} congruence of time-like curves (a family of observers)
in space-time, and there will always exist other families of observers
for whom $\mathcal{H}$ does not block information.
This is obvious since we can always construct a local inertial frame
arbitrarily close to an event on $\mathcal{H}$,
and arrange for time-like curves to cross $\mathcal{H}$.
  
As an explicit illustration,
let us consider a class of space-times with the line element
\begin{equation}
ds^2=\Omega^2(X^a)(-dT^2+dX^2)+dL_\perp^2 ~,
\label{globalmetric}
\end{equation}
where $\Omega(X^a)$ is a non-zero and finite function everywhere
(except possibly at events where the space-time itself has curvature
singularities), and $dL_\perp^2$ vanishes on the $T-X$ plane.
An example of a family of observers with horizon, which we shall repeatedly
come across as a prototype, is provided by a congruence of time-like curves
$X^i (t)=(T(t),X(t),0,0)$:
\begin{equation}
T = x \sinh (at), \quad X = x \cosh (at), \quad a>0,
\label{trajectory}
\end{equation}
where $x$ and $a$ are constants. 
For light rays propagating in the $T-X$ plane, with $ds^2=0$,
the trajectories are lines at $\pm 45^\circ$, just as in flat space-time. 
For all $x>0$, the trajectories in (\ref{trajectory}) are hyperbolas,
confined to the ``right wedge" of the space-time $(\mathcal{R})$
defined by $X>0, |T|<X$.
These observers cannot access any information in the region $|T|>X$,
even when $t \to \infty$, and therefore the null light cone surface
through the origin, $|T|=X$, acts as a horizon for them.
Similarly, Eq.(\ref{trajectory}) with $x<0$ represents a class of observers
accelerating along negative x-axis and confined to the ``left wedge"
of the space-time $(\mathcal{L})$ defined by $ X<0,|T|<|X|$, 
who do not have access to the region $|T|+X>0$.
This example shows that the horizon structure can be ``observer dependent",
and can arise due to the choice of time-like congruence.
  
To avoid possible misunderstanding we stress the following fact:
It is possible to provide \emph{a} definition of horizon in {\it certain}
space-times, e.g. the Schwarzschild space-time, which is purely geometrical.
For example, the boundary of the causal past of the future time-like infinity
in Schwarzschild space-time provides an intrinsic definition of the black hole
event horizon.
But there exist time-like curves,
e.g. those corresponding to observers falling in to the black hole,
for which this horizon does not block information.
The comments made above should be viewed
in the light of whether it is physically relevant and \emph{necessary}
to define horizons as geometric entities,
rather than whether it is \emph{possible} to do so in certain space-times. 
In fact, a purely geometric definition of horizon actually hides certain
physically interesting features.
It is more appropriate to define horizons with respect to a family of
observers (congruence of time-like geodesics) as we have done.
We will follow this point-of-view throughout this paper.
 
To study the physics related to a family of observers, we
 transform to a coordinate system in which these observers are at rest.
One such coordinate system is already provided by Eq.(\ref{trajectory}),
with $(t,x,Y,Z)$ now being interpreted as a new coordinate system,
related to the inertial coordinate system $(T,X,Y,Z)$,
with all the coordinates in the range $(-\infty, \infty)$.
The line interval in these new coordinates is given by 
\begin{equation}
ds^2 =\Omega^2(x^a)( -a^2 x^2 dt^2 + dx^2) + dL_\perp^2 
\label{rindlermetric}
\end{equation}
The light cones $T^2=X^2$ in the $(Y,Z)=$ constant sector,
now correspond to the hypersurface $x=0$ in this new coordinate system.
(It is usually called the Rindler frame, when the space-time is flat
and $\Omega=1$; we shall continue to use this terminology.)
In this coordinate system,
 the $g_{00}=0$ hypersurface---which is just the light cone
through the origin of the original frame---divides the frame
in to two causally disconnected regions.  
The transformation from $(T,X)$ to $(t,x)$, given by Eq.(\ref{trajectory}),
maps the right and left wedges ($\mathcal{R}, \mathcal{L}$)
to ($x>0, x<0)$ regions.
Half of the space-time contained in the future light cone ($\mathcal{F}$)
through the origin ($|X|<T,T>0$), and the past light cone ($\mathcal{P}$)
through the origin ($|X|<T,T<0$) is \emph{not at all} covered by the $(t,x)$
coordinate system.
In other words, both the branches of the light cone $X=+T$ and $X=-T$
collapse to the line $x=0$, and the future and past wedges of the space-time
disappear in this representation.
To recover these missing wedges, we can change the variable from $x$ to
$l=(1/2)ax^2$, casting the line element in the form
\begin{equation}
ds^2 = \Omega^2(x^a)( -2al~dt^2 + \frac{dl^2}{2al}) + dL_\perp^2
\label{standardhorizon}
\end{equation}
and extend the manifold to the range $-\infty<l<\infty$.
(Many examples of horizons we come across in curved space-time have
this structure near the horizon, with $\Omega=1, g_{00}=-1/g_{11}$,
and hence this is a convenient form to use.) 
In the $(t,x)$ coordinates, $t$ is everywhere time-like and the two regions
$x>0$ and $x<0$ are completely disconnected.
In the $(t,l)$ coordinates, $t$ is time-like for $l>0$ and space-like for $l<0$;
the hypersurface $l=0$ acts as a ``one-way membrane",
i.e. signals can go from $l>0$ to $l<0$ but not the other way around.
When we talk of the $x=l=0$ hypersurface as a horizon,
we often have this latter interpretation in mind.
  
The precise definition of a ``horizon" or a ``one-way membrane"
in a general space-time is mathematically intricate, but we do not need it.
There is a fair amount of literature (see for instance, ref.\cite{date})
on the definitions of different classes of horizons
(event, apparent, Killing, dynamic, isolated, etc.),
but fairly simple definitions are adequate for our purpose.
For example, we shall often use the criterion that a boundary defined by
$g_{00}=0,$ $g_{0\alpha}=0$, with all other metric components well behaved
near it, is a one-way membrane.
The fact that such a criterion is not ``intrinsic" or ``geometrical"
is irrelevant for us,
since we have defined the horizon with respect a family of observers,
and will work in a coordinate system in which these observers are at rest.

\subsection{Singular gauges and horizons}

The procedure adopted above started with a globally defined coordinate
system in a manifold (with the metric given by Eq.(\ref{globalmetric})),
and then introduced a congruence of time-like curves which has a horizon.
On transforming to the coordinate system, in which the observers following
this time-like congruence are at rest, we arrived at the metric in 
Eq.(\ref{standardhorizon}), with some components badly behaved at the horizon.
This is the standard geometric language used in general relativity,
where the metric is a tensor living on a manifold
and its specific form depends on the choice of coordinates.

It is, however, possible to provide an alternative point of view and
relate these notions to the gauge degrees of freedom of gravity.
The general freedom of choice of coordinates allows
4 out of 10 components of the metric tensor to be pre-specified.
We take these to be $g_{00}=-N^2,~g_{0\alpha}=N_\alpha$, and use them to
 characterize the observer dependent information.
For example, with the choice
($N=1,N_\alpha=0,g_{\alpha\beta}=\delta_{\alpha\beta}$), the ${\bf x}=$ constant
lines represent a class of inertial observers in flat space-time,
while with ($N=(ax)^2,N_\alpha=0,g_{\alpha\beta}=\delta_{\alpha\beta}$),
the ${\bf x}=$ constant lines represent a family of accelerated observers
with a horizon at $x=0$.
\emph{We only need to change the form of $N$ to make this transition,
whereby a family of time-like trajectories, ${\bf x}=$ constant,
acquires a horizon.} 
Similarly, observers plunging in to a Schwarzschild black hole
will find it natural to describe the metric in the synchronous gauge,
$N=1,N_\alpha=0$ (see for instance, ref.\cite{ll}, \S97),
in which they can access the information present inside the horizon.
On the other hand, the more conservative observers follow the
$r=$constant ($>2M$) lines in the standard foliation with $N^2=(1-2M/r)$,
and the hypersurface $N=0$ acts as the horizon that restricts
flow of information from $r<2M$ to the observers at $r>2M$.

This approach treats the coordinates $x^a$ as markers and $g_{ab}(x)$
as field variables; the gauge transformations of the theory
allow changing the functional forms of $g_{ab}(x)$,
in particular those of $g_{00}(x)$ and $g_{0\alpha}(x)$.
The ${\bf x}$=constant trajectories provide the link between
the gauge choice and the conventional concept of a family of observers.
In this field theoretic language, our fiducial observers can \emph{always}
be taken to follow the time-like trajectories ${\bf x}=$ constant
in a region of space-time, and depending on the gauge choice
they may have access to part or whole of space-time.
The geometric and field theoretic view-points lead to
the same observable results in the classical theory of gravity.
We follow the field theoretic description of gravity in this paper,
particularly because then we can easily describe the contribution of
degrees of freedom beyond the horizon using functional integrals.

Of the gauge variables $N,N_\alpha$, the lapse function $N$
plays a more important role in our discussion than $N_\alpha$,
and we set $N_\alpha=0$ without loss of generality. 
With this choice, the horizon for the ${\bf x}=$ constant family of
observers is specified by the condition $N=0$.
When one changes the gauge used for the description, thereby changing
the form of metric components, a horizon may arise because of the choice
of a \emph{singular gauge} for describing the metric.
This does not happen in case of {\it infinitesimal} gauge transformations,
$\delta g_{ij} = -\nabla_i\xi_j - \nabla_j\xi_i$,
induced by the four gauge functions $\delta x^i=\xi^i$,
since they can only be used to obtain results valid  to first order in $\xi$.
But the  theory is also invariant under finite transformations
which are more ``dangerous".
Of particular importance are the {\it large gauge transformations},
which are capable of changing $N=1$ in the synchronous frame
to a non-trivial function $N(x^a)$ that vanishes along a hypersurface.
In particular, the coordinate transformation $(T,{\bf X})\to(t,{\bf x})$
in the region $|T|<X$ of the Minkowski space-time: 
\begin{equation}
aT=(1+ax)\sinh at ,\quad 1+aX=(1+ax)\cosh at , \quad Y=y, \quad Z=z,
\label{stdrindler}
\end{equation}
changes the metric from $g_{ab}=(-1,1,1,1)$ to $g_{ab}=(-(1+ax)^2,1,1,1)$.
Using such exact large gauge transformations,
we can discuss regions arbitrarily close to the $N=0$ hypersurface,
which here is the Rindler horizon at $x_H=-1/a$.

The neighbourhood of a wide class of horizons can be approximated
by the metric in this gauge, $g_{ab}=(-(1+ax)^2,1,1,1)$.
If the metric is described in a gauge with $g_{\alpha\beta}$
finite and well behaved near the hypersurface $x=x_H$,
then by diagonalizing $g_{\alpha\beta}$ and expanding $N$
in a Taylor series as $N=N'(x_H)(x-x_H)+\ldots$,
we can bring the metric close to the horizon in the Rindler form.
In fact, the transformation (\ref{stdrindler}) generalizes in a simple manner
to even cover some situations where the $N=0$ hypersurface is time dependent
(see for instance, Section 6 of ref.\cite{tpwork1}).

These gauges, which are singular on the horizon,
play an important role in our analysis.
Their significance arises from two principle reasons:

(a) When analytical continuation to Euclidean time is carried out,
the horizon hypersurface reduces to a singular hypersurface,
pinching the space-time and opening up possibilities of interpretation
in terms of topology change \cite{chrisduff,tpinstanton}.

(b) The symmetries of the theory enhance significantly near the
$N=0$ hypersurface.
An interacting scalar field theory, for example, reduces to a $(1+1)$
dimensional CFT near the horizon, and the modes of the scalar field vary as
$\phi_{\omega} \cong \vert x - x_H\vert^{\pm (i\omega / 2a)}$
with $N'(x_H)=a$ (see for instance, Section 3 of \cite{tpwork1}).
Similar conformal invariance occurs for gravitational sector as well
 in a manner similar to  the transformations in case of AdS space-time
(see for instance, ref.\cite{adscft}).
These ideas work because, algebraically, $N\to 0$ makes certain terms
in the diffeomorphisms vanish and increases the symmetry.
There is also a strong indication that most of the results
related to horizons (such as entropy) arise from
the enhanced symmetry of the theory near the $N=0$ hypersurface
(see for instance, ref.\cite{carlip} and references therein). 
(Space-time gauges with $N=0$ hypersurfaces also create difficulties
in canonical quantum gravity using the Wheeler-DeWitt equation;
see for instance, the analysis in ref.\cite{teit} leading to the conclusion:
``\ldots $N$ must be different from zero at all times".)

These issues have surfaced and have been dealt with at different levels of
rigour and correctness in several other works dealing with semiclassical
gravity, but a study of literature shows that it is not easy
to incorporate $N\to 0$ gauges in the analysis of semiclassical gravity.
In our opinion, this is because the standard approach attempts to provide a
\emph{global} theory of quantum/semiclassical gravity, which cannot exist in
a gauge in which a class of observers have only limited access to space-time.
It is necessary to take in to account the role of large gauge transformations
separately, which is what we do in this paper. 
This approach to combining general covariance and quantum theory leads
to as far reaching conclusions as the combination of Lorentz covariance
and quantum theory.
We begin with a discussion of quantum field theory in a singular gauge
(Section III and Appendix A),
and then proceed to study gravity (Sections IV to VI).

\section{Quantum theory in singular gauges} 

To understand the significance of the above discussion
for quantum field theory in curved space-time,
it is instructive to look back at how the special theory of relativity
is combined with quantum dynamics (see ref.\cite{feynman} for instance).
Non-relativistic quantum mechanics has a notion of absolute time $t$ 
and in the path integral representation of non-relativistic quantum mechanics,
one uses only causal paths $X^\alpha(t)$ which ``go forward" in this
absolute time $t$.
This restriction has to be lifted in special relativity, and
the path integral measure has to be changed to sum over paths
$X^a(s)=(X^0(s),X^\alpha(s))$, which go forward in the proper-time $s$,
but either forward or backward in coordinate time $X^0$,
in order to maintain Lorentz invariance.
A particle going backward in time is reinterpreted as an antiparticle
moving forward in time, and U-turns in time direction are identified with
production/annihilation of a particle-antiparticle pair \cite{dirac,feynman}.
The path integral thus sums over paths which could intersect the
$X^0=$ constant hypersurface many times, going forwards and backwards.
The description of such paths is mathematically similar to a system
of an indeterminate number of non-relativistic point particles,
located at different spatial locations at any given time.
Combining special relativity with quantum mechanics \emph{requires}
the use of such constructs with an infinite number of degrees of freedom,
and quantum fields are such constructs
(see for instance, ref.\cite{Padmanabhan:1997qft}).

In the case of a free scalar particle, this result is summarized by:
\begin{eqnarray}
G_F(Y,X)&\equiv&\int_0^\infty ds e^{-ims}\int\mathcal{D}Z^a e^{iA[Y,s;X,0]}
\nonumber\\
&=&\langle 0|T[\phi(Y)\phi(X)]|0\rangle ~.
\label{gf}
\end{eqnarray}
Here $A[Y,s;X,0]$ is the action for the relativistic particle to propagate
from $X^a$ to $Y^a$ in proper time $s$, and the path integral sums over
all paths $Z^a(\tau)$ with these boundary conditions.
The integral over all values of $s$
(with the phase factor $\exp(-iEs)=\exp(-ims)$
corresponding to the energy $E=m$ conjugate to the proper time $s$)
gives the total amplitude for the particle to propagate from $X^a$ to $Y^a$.
There is no notion of a field in this (first) line of the equation,
and the final result is expressed entirely in terms of paths
of virtual and real particles. 
The second line expresses the same quantity in terms of a quantum field.
In either approach, we find that $G_F(Y,X)\neq 0$
when $X^a$ and $Y^a$ are space-like separated;
the propagation amplitude for a relativistic particle to cross
a light cone (or horizon) is \emph{non-zero} in quantum field theory and
 is interpreted
in terms of particle-antiparticle pairs.
There is a well-defined way of ensuring covariance
under Lorentz transformations for this interpretation,
and since all inertial observers see the same light cone structure,
it is possible to construct a Lorentz invariant quantum field theory. 

We still need to ensure that causality is respected in the quantum field theory,
so that information on any $X^0=$ constant hypersurface can be used to
predict the future.
Propagation of all positive energy solutions forward in time,
and all negative energy solutions backward in time,
requires use of retarded propagators and time-ordered products of fields
in Green's functions (as is implicit in second line of Eq.(\ref{gf})).
Analytic continuation to complex values of $t$ is a \emph{must}
for proper implementation of this restriction.
The standard convention is to use the $i\epsilon$-prescription,
which is equivalent to defining the propagator in Euclidean space-time
and analytically continuing it back to Minkowski space-time.
All of this, including the  description in Eq.(\ref{gf}),
is closely tied to the existence of a global time coordinate $T$
(and those obtained by a Lorentz transformation from it)
as well as an invariant light cone structure.
 
The situation becomes more complicated in general relativity,
where general coordinate transformations are allowed and \emph{all families
of observers do not have access to identical regions of space-time}.
In the broadest sense, quantum general covariance will demand
a democratic treatment of all observers,
irrespective of the limited space-time access they may have.
A new conceptual issue comes up while doing quantum field theory in  
such coordinate systems with horizons.
All physically relevant results in the space-time depend on the 
combination $Ndt$, rather than just the coordinate time $dt$.
The Euclidean rotation, $t\to t e^{i\pi/2}$, becomes ambiguous on the
horizon, because the phase of $Ndt$ cannot be defined whenever $N=0$. 
Since it is always possible to choose gauges (coordinate systems)
such that $N=0$ on a hypersurface, we need a new physical principle
to handle quantum field theory in singular gauges.
To resolve this ambiguity, it is necessary to construct a regularization
procedure where singularities of the metric can be avoided.

It is well-known that the $i\epsilon$-prescription used in standard
quantum field theory is equivalent to rotating the time coordinate
to the imaginary axis by $T\to T_E= T e^{i\pi/2}$.
Let us consider what happens to the coordinate transformations 
in Eq.(\ref{trajectory}), and the metric near the horizon,
when such an analytic continuation is performed.
The hyperbolic trajectory in Eq.(\ref{trajectory}) for $x=1/a=$ constant
(for which $t$ measures the proper time), is given in parametric form as
$aT=\sinh at$, $aX= \cosh at$.
This becomes a circle, $aT_E=\sin at_E, aX=\cos at_E$,
with $-\infty<t_E<+\infty$, up on analytical continuation in both $T$ and $t$. 
The  mapping $aT_E=\sin at_E$ is many-to-one
and limits the range of $T_E$  to $|T_E|\le1$.
Eq.(\ref{trajectory}) with $x>0, -\infty<t<\infty$ covers \emph{only}
the right wedge $\mathcal{R}$ of the Lorentzian sector; one needs to take
$x<0, -\infty<t<\infty$ to cover the left wedge $\mathcal{L}$.
Nevertheless, \emph{both wedges} are covered by different ranges of the
``angular" coordinate $t_E$; the range $(-\pi/2)< at_E <(\pi/2)$ covers
$\mathcal{R}$ while the range $(\pi/2)< at_E <(3\pi/2)$ covers $\mathcal{L}$.  
    
The light cones of the inertial frame, $X^2=T^2$,
are mapped to the origin of the $T_E-X$ plane.
The region ``inside" the horizon, $|T|>|X|$,
simply \emph{disappears} in the Euclidean sector.  
Mathematically, Eq.(\ref{trajectory}) shows that $at \to at-i\pi$
changes $X$ to $-X$, i.e. the complex time plane contains information about
the physics beyond the horizon through imaginary values of $t$
(this has been exploited in several derivations of thermality of horizons;
see for instance, refs.\cite{damour,hartle}).
Performing this operation twice shows that $at \to at-2i\pi$ is an identity
transformation, implying $2\pi$ periodicity in the imaginary time $iat=at_E$. 
More generally, all events $\mathcal{P}_n\equiv(t=(2\pi n/a),{\bf x})$,
($n=0,\pm 1,\pm 2, ...$), that correspond to \emph{different} values of $T$,
are mapped to the \emph{same} value of the Euclidean time. 
    
The analytic continuation $t_E=it$ converts the metric near the horizon,
i.e. Eq.(\ref{rindlermetric}) with $\Omega \to 1$ for simplicity, to
\begin{equation}
ds^2 = a^2 x^2 dt_E^2 + dx^2 +dL_\perp^2 
\label{horizonmetric}
\end{equation}
Around the origin of the $t_E-x$ plane,
this is the metric for the surface of a cone.
The phase ambiguity for $N=0$ means that the analytical continuation is
ill-defined at the origin, and a conical singularity is left undetermined.
Indeed, the $i\epsilon$-prescription or the Euclidean continuation
are designed to ensure correct behaviour of the theory only asymptotically,
i.e. as $t\to\pm\infty$.
In case of horizons, we need to regulate the behaviour of the theory
along the whole line $(t\in[-\infty,\infty],x=l=0)$,
and just Euclidean continuation is not able to do that.
We \emph{must} invoke an additional physical principle to regulate horizons.
When the space-time is static (as is the case for the horizons whose
thermodynamical behaviour has been studied), this additional physical
principle is translational invariance along the time direction.
In a static situation,
one cannot single out $t=0$ as the location for a conical singularity,
and the only consistent solution is to have no conical singularity at all.
The Euclidean time $t_E$ thus becomes a genuine angular coordinate with
$0\leq at_E <2\pi$.
  
In our analysis, this procedure of mapping the $N=0$ hypersurface
to a point (origin) in the Euclidean plane plays an important role
in quantum field theory in singular gauges
as well as in the dynamics of gravity (see Section \ref{effectivetheory}). 
To see its role in a broader context, we spell out the steps in our approach:
(i) Consider a family of observers who have a horizon
in a coordinate system in which they are at rest. 
We demand that these observers be able to formulate quantum field theory
entirely in terms of an \emph{effective space-time manifold}
made up of regions accessible to them.
(ii) Since the quantum field theory is well defined only with an analytic
continuation to complex time, it is necessary to construct an effective
space-time manifold \emph{in the Euclidean sector}.
(iii) For a wide class of metrics with horizons, the metric
close to the horizon can be approximated by Eq.(\ref{horizonmetric}).
Analytic continuation to Euclidean time removes the part of the manifold
hidden by the horizon and reduces the horizon to a point (the origin, say).
(iv) The effective manifold for the observers with horizon
can now be thought of as the Euclidean manifold with the origin removed.
This structure is of very general validity, since it only uses
the form of the metric very close to the horizon where it is universal. 
The structure of the metric far away from the origin can be
quite complicated (there could even be another horizon elsewhere),
but the topological structure near the origin is independent of that.
(v) This approach is well defined for static horizons,
and we briefly describe and recover the standard results
of thermal behaviour in Appendix \ref{thermalhorizons}.
 
To handle the possible conical singularity for arbitrary horizons,
we need a more general  regularization procedure.
Even for certain time-dependent but smooth enough horizons,
it may be reasonable to demand the absence of a conical singularity,
and we hope to explore that situation in a future publication.
Here we proceed to discuss the role our paradigm plays
in the dynamics of gravity, in order to obtain new results.
 
\section{The Principle of Effective Theory and Gravity}\label{effectivetheory}

The aspect that different observers may have access to different
regions of space-time and hence differing amount of information,
led us to postulate the ``principle of effective theory''.
In previous sections, we discussed the role played by this principle in
producing the standard results of quantum field theory in curved space-time,
and we shall now discuss the implications of this postulate for gravity.

\subsection{Consequences of the effective theory framework}

As we have already seen, the region inside the horizon disappears 
``in to'' the origin of the Euclidean coordinate system,
for a wide class of horizons.
The principle of effective theory then requires that one should
deal with the theory of gravity in the corresponding effective manifold,
in which the region inaccessible to the family of observers is removed.
(In the example discussed earlier,
the origin of the $X-T$ plane in the Euclidean manifold was removed.) 
When this is done, the information regarding the region inside the
horizon will manifest itself in two different forms:
(i) as periodicity in the imaginary time coordinate and non-trivial
winding number for paths which circle the point that is removed, and
(ii) as a boundary term in the Euclidean action, which has to be defined
carefully taking in to account any contribution from the infinitesimal
region around the point that is removed.
On analytically continuing back to the Lorentzian space-time, the origin
in the Euclidean space-time translates back to the horizon hypersurface.
If we choose to work entirely in the Lorentzian space-time,
we need to take care of the above two effects by: 
(i) restricting the time integration to a finite range
(in defining the action etc.), and
(ii) adding a suitable boundary term to the action describing
contribution of the horizon to gravitational dynamics.
Since the horizon hypersurface is the only common element to the regions
inside and outside, the effect of entanglements across a horizon can
only appear as a boundary term in the action.
It is therefore an inevitable consequence of the principle of effective
theory, that the action functional describing gravity \emph{must} contain
certain boundary terms, which are capable of encoding the information
equivalent to that present beyond the horizon.
This relic of quantum entanglements will survive in the classical limit,
but---being a boundary term---will not affect the equations of motion. 
 
This framework imposes a very strong constraint on the form of action
functional $A_{grav}$ that describes semiclassical gravity.
To study the effects of unobserved degrees of freedom in some space-time
region, let us  divide the space-time manifold in to two regions separated
by a boundary hypersurface.
We choose a coordinate system such that this boundary acts as a horizon for
the observer on one side (say side 1), which usually requires non-trivial
values for the gauge variable $N$. (For example, many space-times with
horizons---Schwarzschild, de Sitter etc.---admit a coordinate chart with
$N^2=g^{rr}$, where $N^2$ has a simple zero at the location of the horizon.)
In quantization of gravity, attempted  through the functional integral
approach based on the action, a sum over all paths with fixed boundary
conditions provides the transition amplitude between initial and final states.
The effective theory for the observer on side 1 is obtained by
integrating out the variables on the inaccessible side (side 2).
With a local Lagrangian, it is formally given by
\begin{equation}
\exp [iA_{\rm eff}(g_1)] \equiv \int [\mathcal{D} g_2] ~
\exp [i(A_{\rm grav}(g_1) + A_{\rm grav}(g_2))] ~.
\end{equation} 
In this integration, the intrinsic geometry of the common boundary has to
remain fixed, e.g. by choosing $g$ to remain continuous across the boundary.
In the semiclassical limit,
integration over $g_2$ can be done by the stationary phase approximation,
and the leading order result is the exponential of the classical action.
The effective theory on side 1 is thus described by the action
$A_{\rm eff}^{\rm WKB}(g_1)$, with 
\begin{equation}
\exp [iA_{\rm eff}^{\rm WKB}(g_1)] =
\exp [i(A_{\rm grav}(g_1) + A_{\rm grav}(g_2^{\rm class}))].
\end{equation}
This result has several non-trivial implications:

(i) Since we expect the effects of unobserved degrees of freedom to be
described by a boundary term, we get the constraint that, when evaluated
on the classical solution, the action $A_{\rm grav}(g_2^{\rm class})$
must be expressible in terms of the boundary geometry.
That is, $A_{\rm grav}(g_2^{\rm class})=A_{\rm boundary}(g_1)$, and
\begin{equation}
\exp [iA_{\rm eff}^{\rm WKB}(g_1)] =
\exp [i(A_{\rm grav}(g_1) + A_{\rm boundary}(g_1))].
\label{thecondition}
\end{equation}
This is a  non-trivial requirement.

(ii) Since the boundary term arises  due to the choice of a specific
coordinate system, in which the boundary acts as a one-way membrane,
$A_{\rm boundary}(g_1)$ will in general depend on the gauge variables
$N,N_\alpha$, and will \emph{not} be invariant under arbitrary
coordinate transformations.
Classically, with the boundary variables held fixed, the equations of
motion remain unaffected by a total divergence boundary term;
the fact that the boundary term is not generally covariant is
unimportant for classical equations of motion.
But even a total divergence boundary term can affect quantum dynamics,
because quantum fluctuations around classical solutions can sense the
properties of the boundary.
We emphasize that the quantum theory is governed by $\exp[iA_{\rm eff}]$
and not by $A_{\rm eff}$.
The boundary term, therefore, will have no effect in the quantum theory,
if the quantum processes keep $\exp[iA_{\rm boundary}]$ single-valued.
This is equivalent to demanding that the boundary term has a discrete
spectrum, with uniform spacing $\Delta A_{\rm boundary}=2\pi$.

(iii) Since the boundary term arises from integrating out the unobserved
degrees of freedom, the boundary term should represent the loss of
information as regards the particular observer (encoded by the choice
of gauge variables $N,~N_\alpha$) and must contribute to the entropy.
This strongly suggests that, in the Euclidean sector, the effective action
$A_{\rm eff}^{\rm WKB}(g_1)$ must have a thermodynamic interpretation.
Indeed, there is a deep connection between the standard thermodynamic
results obtained for quantum fields in curved space-time and the nature
of semiclassical gravity.

\subsection{Form of the boundary action for gravity}

Since our principle puts a strong constraint on the action functional,
it should be possible to deduce the form of the action functional for
gravity from it. We shall now show how this can be done, in two steps.
First we will determine the form of the boundary term, and prove that the
contribution from the horizon must be proportional to the horizon area.
(A general discussion of all possible boundary terms is given
in Appendix \ref{boundaryterms}.)
Then we will determine the complete bulk action using the boundary term
(also see, refs.\cite{tpwork2,tpwork3}). 

The structure of the boundary  terms is easier to understand
in Euclidean space-time with which we shall begin.
Let $\mathcal{M}$ be a compact region
of Euclidean space-time with boundary $\partial\mathcal{M}$.
The geometry of a closed orientable boundary can be fully described
by its outward pointing unit normal, $w^a$, with $w^aw_a=1$.
Since $w^a$ is defined only on the boundary,
the action terms containing it have to be boundary integrals.
Since we would like to have a local Lagrangian description of the theory,
a foliation may be used to extend the value of $w^a$ through the whole
space-time. Such a foliation is by no means unique, and the action
terms must not depend on how the extension of $w^a$ is carried out.
This requirement is guaranteed by demanding that action terms
containing $w^a$ be total divergences; Gauss's law then implies that
$w^a$ defined on the boundary is sufficient to evaluate these terms.
Thus the boundary terms in the effective action for gravity take the form
\begin{equation}
A_{\rm surface}  = \int d^4 x \sqrt{-g}~\nabla_i U^i ~,
\label{firsteqn}
\end{equation}
where we use the notation
$\nabla_i U^i\equiv (-g)^{-1/2}\partial_i[(-g)^{1/2}U^i]$,
irrespective of whether $U^i$ is a genuine four vector or not.
(We use the notation $A_{\rm surface}$ for a general surface term,
and reserve the notation $A_{\rm boundary}$ for the boundary term which
arises after integrating out the metric on one side of a {\it horizon}.)
The symmetry principles of gravity, and the restriction that the action for
the lowest order effective theory should not contain more than two derivatives,
imply that $U^a$ must be constructed from $g_{ij}$, $w_b$ and $\nabla_k$
with the derivative operator acting at most once.
Then the possible candidates for $U^a$ are:
(i) with zero derivatives, $U^a=w^a$;
(ii) with one derivative, 
$U^a=(w^a \nabla_b w^b, w^b \nabla_b w^a, w_b \nabla^a w^b)$.
Of these, $w_b \nabla^a w^b = 0$, due to the normalization $w^b w_b=1$.
Furthermore, the term $\nabla_j (w^b \nabla_b w^j)$ involving the
``acceleration" $a^j= w^b \nabla_b w^j$ integrates to zero,
since the use of Gauss's law converts the integrand to $w_ja^j$
which vanishes identically.
Thus we are left with just two possibilities, $U^a=(w^a,w^a \nabla_b w^b)$.
The most general, lowest order, surface term 
for Euclidean gravity thus takes the form
\begin{equation}
A_{\rm surface}=
\int_{\partial\mathcal{M}} d^3x \sqrt{f}~(\lambda_0+\lambda_1 K_{\rm ext}) ~.
\label{euclsurface1}
\end{equation} 
where $f_{ab}$ is the induced metric on $\partial\mathcal{M}$,
and $K_{\rm ext}=-\nabla_b w^b$ is the Euclidean extrinsic curvature.
The first term is the total volume of the boundary,
while the latter is the integral of the extrinsic curvature
over the boundary.

This expression is quite general.
Now let us consider a space-time with a horizon. 
In the Euclidean sector, there is no light cone,
and the horizon gets mapped to the origin of the $t_E-x$ plane, say.
In the effective manifold, we remove this point, and the surface term
has to be evaluated by a limiting procedure around the origin.
In the region around the origin,
we can take the metric to be approximately Rindler:
\begin{equation}
ds^2_E\approx(\kappa x)^2 dt_E^2+dx^2+dL_\perp^2 ~.
\end{equation}
Consider a surface ${\cal S}_0$,
whose projection on the $t_E-x$ plane is a small circle around the origin.
The first term in Eq.(\ref{euclsurface1})
is proportional to the area of the 3-dimensional boundary.
In the limit of the radius of ${\cal S}_0$ going to zero,
this contribution vanishes.
The interesting contribution comes from the second term.
In absence of a conical singularity, and with $w^i=(0,1,0,0)$, it is
\begin{equation}
A_{\rm boundary}
=-\lambda_1\int d^2x_\perp\int_0^{2\pi/\kappa}dt_E~\partial_x(\kappa x)
=-2\pi \lambda_1{\cal A}_H ~,
\label{euarea}
\end{equation}
where ${\cal A}_H$ is the transverse area of the horizon.
We thus arrive at the conclusion that
\emph{the information blocked by a horizon, and encoded in the surface term,
must be proportional to the area of the horizon}.
Taking note of non-compact horizons, such as the Rindler horizon,
we may state that the entropy (or the information content) per unit area
of the horizon is a constant related to $\lambda_1$.
Writing $\lambda_1\equiv -(1/8\pi \mathcal{A}_P)$, where $\mathcal{A}_P$ is
a fundamental constant with the dimensions of area, the entropy associated
with the horizon becomes $S_H=(1/4)(\mathcal{A}_H/\mathcal{A}_P)$.

We shall next consider the surface terms in the Lorentzian sector.
Since different families of observers have different levels
of access to information, we expect $A_{\rm surface}$ to depend
on the foliation of space-time, and we will choose a particular
$(3+1)$ foliation to determine the form of $A_{\rm surface}$. 
Let $g_{00}=-N^2$, let $u^i=(N^{-1},0,0,0)$ be the four-velocity of observers
corresponding to this foliation (i.e. the normal to the foliation),
and let $a^i=u^j\nabla_ju^i$ be the related acceleration.
Let $K_{ab}=-\nabla_a u_b - u_aa_b$ be the extrinsic curvature
of the foliation, with $K\equiv K^i_i = -\nabla_i u^i$.
(With this standard definition, $K_{ab}$ is purely spatial,
$K_{ab}u^a=K_{ab}u^b=0$; so one can work with the spatial components
$K_{\alpha\beta}$, whenever convenient.)
Once again, we can list all possible vector fields $U^i$
that can appear in (\ref{firsteqn}), and we easily find that
$U^i$ \emph{must} be a linear combination of $u^i$, $u^i K$ and $a^i$.
The corresponding term in the effective action has the form 
\begin{equation}
\label{threedaction}
A_{\rm surface} = \int d^4x~\sqrt{-g}~
\nabla_i \left[ \lambda_0 u^i+\lambda_1 K u^i + \lambda_2 a^i\right] ~,
\end{equation}
where $\lambda$'s are again unknown numerical constants.
We will now compare this with our result in the Euclidean sector
and show that consistency requires $\lambda_1=\lambda_2$.

Let the region of integration be a four volume $\mathcal{V}$,
bounded by two space-like hypersurfaces $\Sigma_1$ and $\Sigma_2$,
and a time-like hypersurface $\mathcal{S}$.
The space-like hypersurfaces are constant time slices with normals $u^i$,
the time-like hypersurface has normal $n^i$, and we choose $n_iu^i=0$.
The induced metric on the space-like hypersurfaces is
$h_{ab} = g_{ab} +u_au_b$, while the induced metric on
the time-like hypersurface is $\gamma_{ab} = g_{ab} -n_an_b$.
The space-like and time-like hypersurfaces intersect on
a two-dimensional surface $\mathcal{Q}$, with the induced metric
$\sigma_{ab} = h_{ab} - n_an_b = g_{ab} +u_au_b -n_an_b$. 
In this foliation, the first two terms of (\ref{threedaction}) contribute
only on the $t=$ constant hypersurfaces ($\Sigma_1$ and $\Sigma_2$),
while the third term contributes only on the time-like surface.
Performing the integrals, we get the surface terms:
\begin{equation}
A_{\rm surface}=\lambda_0\int_{\Sigma} d^3 x~\sqrt{h}
               +\lambda_1\int_{\Sigma} d^3 x~\sqrt{h}K
               +\lambda_2\int_\mathcal{S}dt d^2 x~N\sqrt{\sigma}(n_ia^i) ~.
\label{break}
\end{equation}
Note that in the Euclidean sector we had only one vector field $w^i$,
and hence the $w^ia_i$ type of term, where $a^i$ is the ``acceleration''
corresponding to $w^i$ itself, vanished.
In the Lorentzian sector, we can have a normal $n^i$ of the time-like surface,
which is orthogonal to the normal $u^i$ of the space-like surfaces,
and thus $n_i a^i$ need not vanish. 
When the Lorentzian and Euclidean results are connected by an analytic
continuation, we can express the third term in Eq.(\ref{break})
as suitably defined integrals of extrinsic curvatures of surfaces.
This is achieved using the identity
\begin{equation}
n_i a^i = n_i u^j \nabla_j u^i = -u^j u^i \nabla_j n_i
= (g^{ij} - h^{ij}) \nabla_jn_i =- \Theta + q ~,
\label{thetaink1}
\end{equation} 
where $\Theta\equiv\Theta^a_a$ and $q\equiv q^a_a$ are traces of
the extrinsic curvature of the surface, when treated as embedded
in enveloping manifolds with metrics $g^{ij}$ or $h^{ij}$.
Eq.(\ref{break}) thus becomes,
\begin{eqnarray}
A_{\rm surface}&=&\lambda_0\int_{\Sigma} d^3 x~\sqrt{h} 
               +\lambda_1\int_{\Sigma} d^3 x~\sqrt{h}K
               +\lambda_2\int_\mathcal{S}dt d^2 x~N\sqrt{\sigma}(-\Theta)
               +\lambda_2\int_\mathcal{S}dt d^2 x~N\sqrt{\sigma}q ~,\nonumber\\
	       &=&\lambda_0\int_{\Sigma} d^3 x~\sqrt{h} 
               +\sum_i\lambda_i\int_{\Sigma_i} d^3 x~\sqrt{h}K_{(i)} ~,
\label{breakthru}
\end{eqnarray}
where the sum is over different surfaces,
to which the surface ${\partial\mathcal{M}}$ of Eq.(\ref{euclsurface1})
breaks up on going from Euclidean to Lorentzian sector.
(It is understood that when the normal is embedded in two different manifolds,
one calculates the extrinsic curvature with both the embedding space metrics.)
The different signs for these terms in Eq.(\ref{breakthru}) correctly
encode the distinction between space-like and time-like surfaces
(see for instance, ref.\cite{york}), which is absent in the Euclidean sector.
Comparing this result with Eq.(\ref{euclsurface1}) we see that the terms
involving the extrinsic curvatures match only if $\lambda_1=\lambda_2$.

Let us now specialize to the contribution from a horizon in the Lorentzian
sector (which we treat as the null limit of the time-like surface $\mathcal{S}$,
e.g. the limit $r\to 2M+$ in the Schwarzschild space-time \cite{comment1}).
Here the periodicity of Euclidean time, when translated to Lorentzian
sector, limits the range of time integration, $t\in[0,\beta]$,
and eliminates the surfaces $\Sigma_1$ and $\Sigma_2$.
We get only the contribution from $\nabla_i a^i$ term on the horizon:
\begin{equation}
A_{\rm boundary} = \lambda_2 \int_\mathcal{S} d^4x~\sqrt{-g}~\nabla_ia^i
= \lambda_2 \int_\mathcal{S} dt d^2x~N \sqrt{|\sigma|} (n_\alpha a^\alpha) ~.
\label{boundaryone}
\end{equation} 
As the surface $\mathcal{S}$ approaches the horizon,
the quantity $N (a_in^i)$ tends to $-\kappa$ \cite{comment2},
where $\kappa$ is the surface gravity of the horizon and is constant
over the horizon (see for instance, ref.\cite{surfacegrav}).
Using the relation $\beta\kappa=2\pi$,
the horizon contribution to the action becomes
\begin{equation}
A_{\rm boundary} = -\lambda_2 \kappa \int_0^\beta dt\int d^2x~\sqrt{\sigma}
                 = -2\pi \lambda_2 \mathcal{A}_H ~,
\label{horizonarea}
\end{equation}
where $\mathcal{A}_H$ is the area of the horizon.
This is same as the result in Eq.(\ref{euarea}), since $\lambda_1=\lambda_2$.
It is reassuring to see that the entire formalism works consistently, though
(i) contributions in Euclidean and Lorentzian sectors arise through different
routes, and (ii) there is no light cone structure in the Euclidean sector.

We thus find that the surface term in the Lorentzian sector must have the form
in Eq.(\ref{threedaction}), with $\lambda_1=\lambda_2=-(1/8\pi\mathcal{A}_P)$.
As regards the $\lambda_0$ term, it merely adds a constant to $K$.
The value of this constant is usually fixed by assuming that
if the same foliation is embedded in the flat space-time,
then the surface term should vanish (see for instance, ref.\cite{gibbons}).
We shall ignore it and set $\lambda_0=0$ for the purpose of this paper.
The surface term then has the unique form:
\begin{equation}
A_{\rm surface}=-\frac{1}{8\pi\mathcal{A}_P}\int d^4 x~\sqrt{-g}~
                \nabla_i \left[K u^i + a^i\right] ~.
\end{equation}
In what follows, we choose units such that $\mathcal{A}_P=1$
(in normal units, $\mathcal{A}_P=L^2_{\rm Planck}=G\hbar/c^3$).

\section{The Action Functionals for Gravity}

The complete action functional for gravity contains both bulk and surface terms,
with $L_{\rm grav}=L_{\rm bulk}+\nabla_iU^i$ giving the full Lagrangian.
From the knowledge of $A_{\rm surface}$, it is possible to arrive at
$A_{\rm grav}$ in different ways \cite{tpwork2,tpwork3}.
We essentially need to express the Lagrangian $\nabla_i U^i$ as a
difference between two Lagrangians $L_{grav}$ and $L_{bulk}$, such that:
(a) $L_{grav}$ is a generally invariant scalar,
(b) $L_{bulk}$ is at most quadratic in the time derivatives of the metric
tensor (in lowest order of effective theory), and
(c) $L_{\rm bulk}$ should not explicitly contain total divergences,
since such terms are already accounted for by $L_{\rm surface}$. 
This is just an exercise in differential geometry, and we use the identity,
\begin{equation}
2\nabla_i(Ku^i + a^i) \equiv
R -\left[ {}^{(3)}\mathcal{R} + (K_{ab}K^{ab} - K^2)\right] ~,
\end{equation}
where $R$ and ${}^{(3)}\mathcal{R}$ are the scalar curvatures of the four
dimensional space-time and the $t=$ constant hypersurface respectively.
(This result is mentioned in ref.\cite{mtw}, p.520, Eq.(21.88);
a simple derivation is given in Appendix \ref{actionstructure}).
This allows us to identify the action functional for gravity as
the Einstein-Hilbert action
\begin{equation}
A_{\rm EH} \equiv {1\over 16\pi} \int d^4x~\sqrt{-g}~R ~,
\end{equation}  
with the decomposition
\begin{equation}
R \equiv L_{\rm EH}
= L_{\rm ADM} - 2\nabla_i(Ku^i + a^i) \equiv L_{\rm ADM}+L_{\rm div} ~.
\label{ehandadm}
\end{equation}
Here
\begin{equation}
L_{\rm ADM} = {}^{(3)}\mathcal{R} + (K_{ab}K^{ab} - K^2)
\label{defadml}
\end{equation}
is the ADM Lagrangian \cite{ADM} quadratic in $\dot g_{\alpha\beta}$,
and $L_{\rm div} = -2\nabla_i(Ku^i + a^i)$ is a total divergence.
Neither $L_{\rm ADM}$ nor $L_{\rm div}$ is generally invariant.
For example, $u^i$ explicitly depends on $N$,
which changes when one makes a coordinate transformation
from the synchronous frame to a frame with $N\neq 1$.
(The most general effective action for gravity also contains a
cosmological constant term, obtained by adding a real constant to $R$.
This term is not important in our analysis, and we leave it out.)

The arguments given in Section \ref{effectivetheory}
show that only a very special kind of action $A_{\rm grav}$ can fit
the structure described by Eq.(\ref{thecondition}) in a natural fashion.
Incredibly enough, the conventional action used for gravity
fits the bill, though it was never introduced in this light.
The work in \cite{tpwork2,tpwork3} as well as the discussion in
Appendix \ref{boundaryterms} show that this is because of
the strong constraints imposed on the thermodynamic route to gravity.
We now explore this aspect.

We start by noting that there is a conceptual difference between the
$\nabla_i(Ku^i)$ term and the $\nabla_i a^i$ term that occur in $L_{\rm div}$.
This is obvious in the standard foliation, where $Ku^i$ contributes
on the constant time hypersurfaces, while $a^i$ contributes on the
time-like or null surfaces that separate the space-time in to two regions
(as in the case of a horizon).
To take care of the $Ku^i$ term more formally, we recall that
the form of the Lagrangian used in functional integrals depends
on the nature of the transition amplitude one is interested in computing,
and one is free to choose a suitable perspective.
For example, in non-relativistic quantum mechanics, if one uses the
coordinate representation, the probability amplitude for the dynamical
variables to change from $q_1 $ (at $t_1$) to $q_2$ (at $t_2$) is given by
\begin{equation}
\psi(q_2,t_2) = \int dq_1 K \left( q_2,t_2;q_1,t_1 \right) \psi(q_1,t_1) ~,
\end{equation}
\begin{equation}
K \left( q_2,t_2;q_1,t_1 \right) = \sum\limits_{\rm paths}
\exp \left[ {i\over\hbar} \int dt~L_q(q,\dot q) \right] ~,
\label{qsopa}
\end{equation}
where the sum is over all paths connecting $(q_1,t_1)$ and $(q_2,t_2)$,
and the Lagrangian $L_q(q,\dot q)$ depends on $(q,\dot q)$. 
It is, however, quite possible to study the same system in momentum space,
and enquire about the amplitude for the system
to have a momentum $p_1$ at $t_1$ and $p_2$ at $t_2$. 
From the standard rules of quantum theory, the amplitude for the particle
to go from $(p_1,t_1)$ to $(p_2,t_2)$ is given by the Fourier transform
\begin{equation}
G \left( p_2,t_2;p_1,t_1 \right) \equiv \int dq_2 dq_1
~K \left( q_2,t_2;q_1,t_1 \right)  
~\exp \left[ -{i\over\hbar} \left( p_2 q_2 - p_1 q_1 \right) \right] ~.
\label{qftofq} 
\end{equation}
Using (\ref{qsopa}) in (\ref{qftofq}), we get
\begin{eqnarray}
G \left( p_2,t_2;p_1,t_1 \right) &=& \sum\limits_{\rm paths}
\int dq_1 dq_2 \exp \left[ {i\over\hbar}
\left\{ \int dt~L_q - \left( p_2 q_2 - p_1 q_1 \right) \right\} \right]
\nonumber \\
&=& \sum\limits_{\rm paths} \int dq_1 dq_2 \exp \left[ {i\over\hbar}
\int dt \left\{ L_q - {d \over dt} \left( pq \right) \right \} \right]
\nonumber \\
&=& \sum_{\rm paths} {} \exp \left[ {i\over\hbar}
\int L_p(q, \dot q, \ddot q)~dt \right] ~.
\label{lp}
\end{eqnarray}
In arriving at the last expression, we have
(i) redefined the sum over paths to include integration over $q_1$ and $q_2$,
and (ii) upgraded the status of $p$ from the role of a parameter in the
Fourier transform to the physical momentum $p(t)=\partial L/\partial \dot q$.
This result shows that, given any Lagrangian $L_q(q,\partial q)$
involving only up to the first derivatives of the dynamical variables,
it is \emph{always} possible to construct another Lagrangian
$L_p(q,\partial q,\partial^2q)$ involving up to second derivatives,
such that it describes the same dynamics but with different boundary
conditions \cite{tpwork2}.
The prescription is:
\begin{equation}
L_p = L_q - {d\over dt} \left( q{\partial L_q \over \partial\dot q} \right) ~.
\label{lbtp}
\end{equation}
When using $L_p$, one keeps the \emph{momenta} $p'$s fixed
at the end points rather than the \emph{coordinates} $q'$s.
This boundary condition is specified by the subscripts on the Lagrangians.

This result generalizes directly to multi-component fields.
If $q_A(x^i)$ denotes a component of a field (which could be a component
of a metric tensor $g_{ab}$, with $A$ formally denoting pairs of indices),
then we just need to sum over $A$.
Since $L_{ADM}$ is quadratic in $\dot g_{\alpha\beta}$, we can treat
$g_{\alpha\beta}$ as coordinates and obtain another Lagrangian $L_\pi$
in the momentum representation.
The canonical momentum corresponding to $q_A=g_{\alpha\beta}$ is
\begin{equation}
p^A = \pi^{\alpha\beta}
= \frac{\partial (\sqrt{-g}~L_{ADM})}{\partial \dot g_{\alpha\beta}}
=- \sqrt{-g} {1 \over N} (K^{\alpha\beta}- g^{\alpha\beta}K) ~,
\end{equation}
so that the term $d(q_Ap^A)/dt$ is just the time derivative of
\begin{equation}
g_{\alpha\beta}\pi^{\alpha\beta}
= -\sqrt{-g} {1 \over N}(K-3K) = \sqrt{-g} (2Ku^0) ~.
\end{equation}
Since
\begin{equation}
{\partial \over \partial t} (\sqrt{-g}~Ku^0)
= \partial_i (\sqrt{-g}~Ku^i) = \sqrt{-g}~\nabla_i (Ku^i) ~,
\end{equation}
the combination
$\sqrt{-g}~L_\pi \equiv \sqrt{-g} [L_{\rm ADM} - 2 \nabla_i(Ku^i)]$
describes the same system in the momentum representation with
$\pi^{\alpha\beta}$ held fixed at the end points \cite{york}.
Switching over to this momentum representation, the relation between the
action functionals in Eq.(\ref{ehandadm}) can now be expressed as 
\begin{equation}
A_{\rm EH} = A_\pi + A_{\rm boundary} ~,
\end{equation}
where
\begin{equation}
A_\pi \equiv A_{\rm ADM} - {1 \over 8\pi}\int\sqrt{-g}~d^4x~\nabla_i(Ku^i) ~,
\label{defmomspace}
\end{equation}
is the ADM action in the momentum representation, and 
\begin{equation}
A_{\rm boundary} =- {1 \over 8\pi} \int d^4x~\sqrt{-g}~\nabla_ia^i
= -{1 \over 8\pi} \int dt \int_\mathcal{S} d^2x~N \sqrt{\sigma}
(n_\alpha a^\alpha) 
\label{boundary}
\end{equation} 
is the boundary term arising from the integral over the horizon hypersurface.

\section {Semiclassical Quantization of Gravity}

In case of space-times without boundary, it does not matter if one works
with $A_{\rm EH}$ or $A_{\rm ADM}$ or $A_\pi$, since they all differ from
each other by total divergences.
On the contrary, while dealing with space-times with boundaries,
it is crucial to use an appropriate action in the functional integral
for gravity.
We believe that the correct action to use in the functional integral is
$A_{\rm ADM}$ or equivalently $A_\pi$ (which describes the same system
in the momentum representation), since it is quadratic
in the time derivatives of the true dynamical variables $g_{\alpha\beta}$. 

To study the effects of unobserved degrees of freedom in some space-time region,
let us divide the space-time manifold in to two regions with a boundary
hypersurface separating them, and choose a coordinate system such that
this boundary acts as a horizon for the observer on one side (side 1, say).
The effective theory for this observer is obtained by integrating out the
variables on the inaccessible side (side 2):
\begin{equation}
\exp [iA_{\rm eff}(g_1)] \equiv \int [\mathcal{D} g_2]
~\exp [i(A_{\pi} (g_1) + A_{\pi}(g_2))] ~.
\end{equation} 
We have chosen $A_\pi$ to be the action functional describing gravity,
so the functional integral is to be evaluated holding the extrinsic
curvature of the boundary fixed.
In the absence of any matter,
we have $R=0$ for the classical solution of gravity.
The semiclassical action then becomes, $A_{\rm EH}^{\rm WKB}=0$, and
\begin{equation}
A_{\pi} (g_2^{\rm WKB}) = - A_{\rm boundary}^{n_2}(g_2)
= A_{\rm boundary}^{n_1}(g_1)
= -{1 \over 8\pi} \int d^4x~\sqrt{-g}~\nabla_ia^i ~,
\label{divdef}
\end{equation}
where $n_1=-n_2$ denote the outward normals of the two sides.
(As an aside, we mention that if the matter on side 2 is described by a
scale invariant action, making the energy-momentum tensor traceless,
$T=0$, then the same result holds.
If the matter has non-zero $T$, then we get an extra phase factor involving
the volume integral of $T$ but independent of the boundary degrees of freedom.
This phase factor does not affect our conclusions, since we are only
concerned with phases which \emph{change} under coordinate transformations.
We are not including matter degrees of freedom in our discussion here,
and hope to address them in a future publication.)
Thus, in the semiclassical limit, we indeed obtain
a boundary term involving the gravitational degrees of freedom,
as anticipated by the principle of effective theory. 
At the next order in the semiclassical approximation,
the effective theory on side 1 is described by the action 
\begin{equation}
\exp [iA_{\rm eff}^{\rm WKB}(g_1)]
= \exp [i(A_{\pi}(g_1) + A_{\rm boundary}^{n_1} (g_1))] \times {\rm det}(Q) ~.
\end{equation}
Here det$(Q)$ arises from integration over quantum fluctuations,
and we ignore it in the lowest order analysis.

The extra term, $A_{\rm boundary}$, is a total divergence and does not
change the equations of motion for side 1 in the classical limit.
But it can affect the quantum theory, unless
$2\sqrt{-g}~\nabla_ia^i = 2 \partial_\alpha (\sqrt{-g}~a^\alpha)$
does not contribute. In certain situations this term vanishes:
(a) If one uses a synchronous coordinate system with $N=1,~N_{\alpha}=0$,
in which there is no horizon.
In the case of a black hole space-time, for example, this coordinate system
will be used by a family of in-falling observers.
(b) If the integration limits for $2 \partial_\alpha (\sqrt{-g}~a^\alpha)$
could be taken at, say origin and spatial infinity, where the contribution
actually vanishes due to $N_\alpha\to 0,~N\to$ constant at these limits.

For a generic observer, however, we cannot ignore the contribution of this
term, and we have to deal with the boundary action (\ref{boundary}).
More explicitly, if we compare the synchronous frame (for which $N=1$),
with the one obtained by the infinitesimal gauge transformation given by
\begin{equation}
\xi^\alpha = \int dt~N^2 g^{\alpha\beta}
              \frac{\partial\xi^0}{\partial x^\beta} + f^\alpha(x^\beta) ~.
\label{largegauge}
\end{equation}
we find that $N_\alpha=0$ continues to hold but $N$ changes
(see ref.\cite{ll}, \S97); hence the value of the boundary term can change.
Our principle of effective theory requires that this coordinate/observer
dependent term should not affect the quantum amplitudes.
The only way to ensure this is to make $\exp[iA_{\rm boundary}]$
single-valued, i.e. demand that 
\begin{equation}
A_{\rm boundary} = 2\pi n + {\rm constant},~ n={\rm integer}.
\end{equation}
Then $\exp[iA_{\rm boundary}]$ becomes an overall phase, and the physics
on side 1 is determined by $A_\pi(g_1)$ as originally postulated.
The values of $A_{\rm boundary}$ measured from side 1 and side 2 are of
opposite sign, because of the opposite direction of their outward normals.
The spectrum of $A_{\rm boundary}$ is therefore symmetric about zero:
\begin{equation}
A_{\rm boundary} = 2\pi m ~,
\label{spectrum}
\end{equation}
with two possible sequences for $m$, either $m \in \{0,\pm1,\pm2,\ldots\}$
or $m \in \{\pm1/2,\pm3/2,\ldots\}$.
The boundary term---which is not generally invariant---may be different
for different observers, but the corresponding quantum operators need
not commute either, thereby eliminating any possible contradiction.  
(This is analogous to the fact that, in quantum mechanics,
the component of angular momentum measured along \emph{any} axis
is quantized, irrespective of the orientation of the axis.)
The action with a uniformly spaced spectrum, $A=2\pi m\hbar$,
has a long and respectable history in quantum field theories,
and our analysis gives a well defined realization of this property
for the semiclassical limit of quantum gravity.

As we have said several times now, a static horizon leads to an effective
Euclidean manifold with a point removed and periodic Euclidean time.
In the Lorentzian sector this implies that, for any static horizon, 
we must take the range of time integration to be $[0,\beta=2\pi/\kappa]$
where $\kappa$ is the surface gravity.
Hence the quantization condition for space-times with static horizons becomes 
\begin{equation}
A_{\rm boundary} = -{1 \over 8\pi} \int_0^\beta dt
\int d^2x~N \sqrt{\sigma}~(n_\alpha a^\alpha) = 2\pi m ~,
\label{condition}
\end {equation} 
with $m\ge0$ for the observer outside the one-way horizon.
This leads to he following consequences:

(i) We have seen that in all static space-times with horizons,
this boundary term is proportional to the area of the horizon,
\begin{equation}
A_{\rm boundary}=\frac{1}{4}({\rm Horizon~Area})
\equiv\frac{1}{4}\mathcal{A}_H ~.
\label{contribution}
\end {equation}
Our result therefore implies that the area of the horizon,
as measured by \emph{any} observer blocked by that horizon, will be quantized.
(In normal units, $A_{\rm boundary}=2\pi m\hbar$ and
$\mathcal{A}_H=8\pi mL_{\rm Planck}^2=8\pi m(G\hbar/c^3)$.)
In particular, any flat spatial surface in Minkowski space-time can be made
a horizon for a suitable Rindler observer, and hence all area elements
in even flat space-time must be intrinsically quantized.
In the quantum theory, the area operator for one observer
need not commute with the area operator of another observer,
and there is no inconsistency in all observers measuring quantized areas.
The changes in area, as measured by any observer, are also quantized,
and the minimum detectable change is of the order of $L_{\rm Planck}^2$.
It can be shown, from very general considerations, that there is an
operational limitation in measuring areas smaller than $L_{\rm Planck}^2$,
when the principles of quantum theory and gravity are combined \cite{tplimit};
our result is consistent with this general analysis.
Quantization of areas also arises in loop quantum gravity 
(see for instance, ref.\cite{Rovelli:1998gg} and references therein).

While there is considerable amount of literature suggesting that
the area of \emph{a black hole horizon} is quantized
(for a small sample, see refs.\cite{areaquant}),
we are not aware of any result which is as general as suggested above
or derived so simply.
Even in the case of a Schwarzschild black hole
all the results in the literature do not match in detail,
nor is it conceptually easy to relate them to one another.
For example, a simple procedure to derive the area spectrum of a Schwarzschild
black hole from canonical quantum gravity is to proceed as follows. 
Classically, the outside region of any spherically symmetric collapsing
matter of finite support is described by the Schwarzschild metric,
with a single parameter $M$.
Since the pressure vanishes on the surface of the collapsing matter,
any particle located on the surface will follow a time-like geodesic
trajectory $a(t)$ in the Schwarzschild space-time.
Because of the extreme symmetry of the model as well as the constancy of $M$,
the dynamics of the system  can be mapped to the trajectory of this particle.
The action describing this trajectory can be taken to be 
\begin{equation}
A = \int dt~(p\dot{a} - H(p,a)) ~,~~
H = \frac{1}{2} \left( \frac{p^2}{a} + a \right) ~,
\label{hamqc}
\end{equation}
which is precisely the action that arises in the study of closed
dust-filled Friedmann models in quantum cosmology.
(It is possible to introduce a canonical transformation from $(p,a)$ to another
set of variables $(P,M)$ such that the Lagrangian becomes $L= P \dot M - M$.
This is similar to the set of variables introduced by Kucha\v{r} \cite{kuchar}
in his analysis of spherically symmetric vacuum space-times.)
This is expected since---classically---the Schwarzschild exterior
can be matched to a homogeneous collapsing dust ball,
which is just the Friedmann universe as the interior solution.
Now, it is obvious that there is {\it no} unique quantum theory for the
Hamiltonian in (\ref{hamqc}) because of  operator ordering problems.
However, several sensible ways of constructing a quantum theory
from this Hamiltonian lead to discrete area spectrum
as well as a lower bound to the area (see for example \cite{tpqc}). 
There are many variations on this theme as far as the area of a black hole
horizon is concerned, but we believe our approach is conceptually simpler
and bypasses many of the problems faced in other analyses. 
  
(ii) The boundary term originated from our integration of the unobserved
gravitational degrees of freedom hidden by the horizon.
Such an integration should naturally lead to the entropy of the unobserved
region, and we get---\emph{as a result}---that the entropy of a horizon is
always one quarter of its area.
Our analysis also clarifies that this horizon entropy is the contribution
of quantum entanglement across the horizon
of the gravitational degrees of freedom.
Further it makes \emph{no} distinction between different types of horizons,
e.g. the Rindler and Schwarzschild horizons.
In contrast to earlier works, most of the recent works---especially the ones
based on CFT near horizon---do not make any distinction between different
types of horizons.
We believe all horizons contribute an entropy proportional to area for
observers whose vision is limited by those horizons.
As we stressed before, the entropy of the black hole is also observer dependent,
and freely falling observers plunging in to the black hole will not attribute
any entropy to the black hole.
More generally, the analysis suggests a remarkably simple,
thermodynamical interpretation of semiclassical gravity.
This equivalence is explored in detail for spherically symmetric
space-times in \cite{tpcqg}.

The crucial feature which we have exploited is that the conventional action
for gravity contains a boundary term involving the integral of the normal
component of the acceleration.
As far as we know,
this term has not been brought to center stage in any of the previous analyses
(a possible reason is mentioned in Appendix \ref{actionstructure}). 
Since the actions for matter and gravity will be additive
in the full theory of gravity, the boundary term also has implications
for the dynamics of the horizon in the full theory.
This in turn will require handling a horizon which is time dependent,
in the sense that $N=0$ on the surface $x=x_H(t)$
in some suitably chosen coordinate system.
We hope to address this problem in a future publication.  

\section{Summary and Outlook}

Gravity is a geometric theory of space-time, and its natural setting is in
terms of the metric tensor as a function of the coordinates, $g_{ab}(x)$.
The fact that there is a maximum speed for propagation of any physical
signal, i.e. the speed of light, means that no observer can access phenomena
occurring outside his light cone.
In any space-time, there exist choices of time-like congruences
(i.e. family of observers), whose global light cone structure
makes part of the space-time inaccessible to them.
The natural coordinate system for such observers is
equivalent to describing the metric in a singular gauge.
A realistic physical theory must be formulated in terms of
whatever variables the observer has access to;
contribution of the unobservable regions must be such that it can be
re-expressed in terms of the accessible variables.
This principle of effective theory provides a powerful constraint on the
theory of gravity, when it is demanded that the same formulation of the
theory should be used by all observers, irrespective of whether or not
their coordinate choice blocks their access to some regions of space-time.

An effective theory for any observer is constructed by integrating
out the degrees of freedom inaccessible to him.
Effective theories are most predictive when they contain only a small number
of terms, because the coefficients of these terms are empirical parameters.
For this purpose, possible terms in the effective theory action are
restricted using symmetry principles and truncated to low orders in the
derivative expansion.
The possible terms that may appear in the effective action for gravity can be
obtained from general principles, as described in Appendix \ref{boundaryterms}.
Apart from the usual bulk terms of general relativity,
all the additional terms that appear are integrals of total divergences.
When the boundary of space-time is a null surface such as the horizon,
only a unique boundary term survives in the effective action;
in a sense, the principle of effective theory leads to holographic
behaviour for one-way membranes.
It should be noted that the effective theory description will be valid
for any extension of general relativity (supergravity, string theory,
loop quantum gravity, anything else).
Some of the extensions may allow topological changes of space-time,
and the effective theory analysis is fully capable of tackling them.

To make any quantum field theory respect causality,
it is mandatory to analytically extend it to complex time plane.
In the Euclidean sector of a space-time, any null surface reduces
to a point, and the resultant conical singularity has to be regularized.
For a static horizon, this regularization amounts to elimination
of the conical singularity, and that explicitly shows that
the boundary term is proportional to the horizon area.
In fact, the complete bulk action for gravity can be constructed from
this boundary term, using nothing but the symmetry principles.

We have used the ADM formulation of gravity to study the consequences
of the boundary term on the dynamical evolution of space-time.
With a 3+1 foliation covering the whole space-time, the boundary term
can be explicitly obtained by integrating out the inaccessible degrees
of freedom beyond the horizon in the semiclassical approximation.
To protect the quantum amplitudes from acausal entanglements across
horizons, the boundary term has to satisfy $\exp(iA_{\rm boundary})=1$.
The quantization condition (\ref{condition}) fixes the normalization
of the boundary term, and produces a uniformly spaced spectrum for it.

It is worth observing that even though the total divergence form of
$A_{\rm boundary}$ and its quantization (\ref{spectrum}) would hold in
the complete quantum theory of gravity, the interpretation of
$A_{\rm boundary}$ in terms of the horizon area holds only in the
lowest order effective theory, and in the semiclassical limit.
Higher order corrections can change the form of $A_{\rm boundary}$
so that it no longer is proportional to the horizon area,
while the true quantum area operator can differ from the
$A_{\rm boundary}$ term which only measures the projection of
the area operator on the horizon surface.
Staying within the lowest order effective theory means that one should
not go very close to the horizon, and semiclassical limit means that
the horizon area parameter $m$ should be large enough.
With the usual power counting counting arguments, these conditions can
be quantified to mean that our result for $A_{\rm boundary}$ is valid
up to $\mathcal{O}(L^{-1}_{\rm Planck})$ and $\mathcal{O}(\ln m)$ corrections.

Since the boundary term arises from integrating out the inaccessible
degrees of freedom, it is natural to connect it to the entropy of
the region blocked by the horizon.
Analytical continuation of the boundary term to Euclidean time
confirms this expectation, and various thermodynamic relations
describing properties of the horizon follow.
In our framework, the horizon entropy arises purely from integrating out
the gravitational degrees of freedom, and it is highly tempting to interpret
the discrete value of the boundary term as the result of quantum topological
changes of the region hidden by the horizon.
To really discover the quantum topological features of gravity,
we need to go beyond the framework presented here,
and that is under investigation \cite{inprogress}.

\section*{Acknowledgments}

We thank A.P. Balachandran, J. Bicak, G. Date, J.Katz, J.Oppenheim,
K. Subramanian, Tulsi Dass and U. Yajnik
for comments on the earlier version of this paper.

\appendix
\section{Propagators in Singular Gauges and Thermality of Horizons
\label{thermalhorizons}}
 
Let us begin by computing the amplitude for a particle to propagate
from an event $P$ to another event $P'$ with an energy $E$ \cite{hartle}.
From the general principles of quantum mechanics,
this is given by the Fourier transform of the Green's function
$G_F[P \to P']$ with respect to the time coordinate.
The vital factor, of course, is which time coordinate is used as 
the conjugate variable to energy $E$.
Consider, for example, the flat space-time situation
in the \emph{Rindler coordinates} ($t,l,0,0$),
with $P'$ being some point on the $T=t=0$ axis in $\mathcal{R}$,
and $P$ being some event in $\mathcal{F}$.
The amplitude $G_F[P \to\mathcal P']$ will now correspond to
a particle propagating from the inside of the horizon to the outside.
(The fact that this is non-zero in quantum field theory
is a necessary condition for the rest of the argument.)
When the energy $E$ is measured with respect to the
\emph{Rindler time coordinate}, this amplitude is given by
\begin{equation}
\mathcal{A}(E;P \to P') = \int_{-\infty}^\infty dt~e^{-iEt}
  G_F[P(t,{\bf y}) \to P'(0,{\bf x})]
\end{equation}
(The minus sign in $\exp(-iEt)$ is due to the fact that
$t$ is the time coordinate of the initial event $P$.)
Shifting the integration by $t\to t-i(\pi/a)$, we pick up a pre-factor
$\exp(-\pi E/a)$; furthermore, the event $P$ becomes the event $P_R$
obtained by reflection in the origin of the inertial coordinates.
We thus obtain
\begin{eqnarray}
\mathcal{A}(E;P \to P') &=& e^{-\pi E/a}
  \int_{-\infty}^\infty dt~e^{-iEt} G_F[P_R \to P'] \nonumber\\
&=& e^{-\pi E/a} \mathcal{A} (E;P_R \to P') ~.
\end{eqnarray}
The reflected event $P_R$ is in the region $\mathcal{P}$;
the amplitude $\mathcal{A} (E ;P_R \to P')$ corresponds to the emission
of a particle across the past horizon (``white hole'' in the case of
Schwarzschild space-time) in to the region $\mathcal{R}$.
By time reversal invariance, the corresponding probability is the same
as the probability $P_{\rm abs}$ for the black hole to absorb a particle.
It follows that the probabilities for emission and absorption of a particle
with energy $E$ across the horizon are related by 
\begin{equation}
  P_{\rm em} = P_{\rm abs} \exp \left(-\frac{2\pi E}{a}\right) ~.
\end{equation}
This result can be generalized to any other horizon,
since the ingredients which we have used are common to all of them.
The translation in time coordinate, $t\to t-i(\pi/a)$,
requires analyticity in a strip of width ($\pi/a$) in the complex time plane,
which can be proved in quite general terms. 
    
The fact that the propagation amplitudes between two events in flat
space-time can bear an exponential relationship is quite unusual.
The crucial feature to note is that the relevant amplitude is defined
at constant energy $E$, which in turn involves Fourier transform of the
Green's function with respect to the Rindler time coordinate $t$.
It is this feature which leads to the exponential factor in virtually
every derivation of this effect.
  
To see this result more explicitly, let us ask how the amplitude in
Eq.(\ref{gf}), in flat space-time, will be viewed by observers following
the trajectories in Eq.(\ref{trajectory}) for $x=1/a=$ constant.
For mathematical simplicity, let us consider a massless particle,
for which $G_F(Y,X)=- (4\pi^2)^{-1} [s^2(Y,X)- i\epsilon]^{-1}$ where
$s^2(Y,X)$ is the invariant space-time interval between the two events.
Treating $G_F(Y,X)$ as a scalar, we find that
\begin{equation}
G_{\rm F}(Y(t),X(t')) = -\frac{1}{4\pi^2}
  \frac{(a/2)^2}{{\rm sinh}^{2} \left[a(t-t')/2\right] - i\epsilon} ~.
\label{scalaramp}
\end{equation}
The first striking feature of this amplitude is that
it is periodic in imaginary time under the change  $it\to it+2\pi/a$,
which arises from the fact that Eq.(\ref{trajectory}) has this property.
In the limit $a\to 0$ (i.e. going far away from the horizon),
$G_F$ is proportional to $[(t-t')^2 -i\epsilon]^{-1}$
which is the usual result in inertial coordinates. 
Next, using the series expansion for ${\rm cosech}^2z$,
the propagator in Eq.(\ref{scalaramp}) can be expressed as a series:
\begin{equation}
G_F(\tau) = -\frac{1}{4\pi^2} \sum_{n=-\infty}^{n=\infty} 
  \left[\left(\tau+2\pi i n a^{-1}\right)^{2} - i\epsilon\right]^{-1} ~,
\label{trouble}
\end{equation}
where $\tau=(t-t')$.
The $n=0$ term corresponds to the inertial propagator (for $a=0$),
and the other terms describe new effects.
If we interpret the Fourier transform of $G_F(\tau)$ as the amplitude for
propagation in energy space, Eq.(\ref{trouble}) will give an amplitude
\begin{equation}
\Delta G(|E|)
  \equiv \int_{-\infty}^{+\infty} d\tau~e^{iE\tau } \Delta G(\tau) 
  = \frac{1}{2\pi} \frac{|E|}{\exp(2\pi|E|/a) -1} ~,
\label{planckone}
\end{equation}
where $\Delta$ indicates that the $n=0$ term has been dropped.
  
The new feature that has come about is the following:
In computing $G_F(Y(t),X(t'))$ using Eq.(\ref{gf}),
we sum over paths which traverse all over the $X-T$ plane even though
the two events $Y(t)$ and $X(t')$ are in the right wedge.
The paths that contribute to Eq.(\ref{gf}) can criss-cross the horizon
several times even though the region beyond the horizon is inaccessible
to the observers following the trajectories in Eq.(\ref{trajectory}).
The net effect of paths criss-crossing the horizon is the extra term
in Eq.(\ref{planckone}).
In fact, the $n>0$ terms in Eq.(\ref{trouble}) contribute for $E<0$,
and the $n<0$ terms contribute for $E>0$.
The result in Eq.(\ref{planckone}) also shows that
$\Delta G(|E|)/\Delta G(-|E|)=\exp(-2\pi |E|/a)$,
which can be interpreted as the ratio of probabilities
for a particle to cross the horizon in opposite directions.
  
These features emerge more dramatically in the Euclidean sector
\cite{chrisduff,vandam,tpinstanton}.
The Euclidean Green's function is $G_E\propto R^{-2}$,
where $R^2$ is the Euclidean distance between the two events.
To express the Euclidean Green's function in terms of $t$ and $t'$,
we need to analytically continue in $t$ by $t\to t_E= it$.
In terms of $t_E,t_E'$, the Green's function becomes,
\begin{equation}
G_{\rm E}(Y_E(t_E),X_E(t_E')) = \frac{1}{4\pi^2}
  \frac{(a/2)^2}{\sin^{2}\left[a(t_E-t_E')/2\right]} ~,
\end{equation}
and can be expressed as a series:   
\begin{equation}
G_E(t_E-t_E') = \left(\frac{a^2}{4\pi^2}\right)
  \sum_{n=-\infty}^{n=\infty} \left[\theta-\theta'+ 2\pi n\right]^{-2} ~,
\label{troublee}
\end{equation}
with $\theta\equiv at_E$.
Clearly, each term in the sum can be interpreted as due to a loop
that winds $n$ times around the circle of radius $x=1/a$
in the $\theta$ direction.  
These winding paths go over the $X<0$ region of Minkowski space, and so
the virtual paths which wind around the origin in the Euclidean sector
contain information about the region beyond the horizon even though $x>0$.
There is thus an intimate connection between thermal behaviour for the
Green's functions, and removal of origin from the Euclidean sector
that leads to winding number for paths.
More details can be found in the literature cited above.
   
\section{The Structure of the Action Functional\label{actionstructure}}

We foliate the space-time by a series of space-like hypersurfaces $\Sigma$
with normals $u^i$. From the relation
$R_{abcd}u^d = (\nabla_a \nabla_b - \nabla_b \nabla_a) u_c$,
we obtain
\begin{eqnarray}
R_{bd} u^b u^d &=& g^{ac} R_{abcd} u^b u^d
 = u^b \nabla_a \nabla_b u^a - u^b \nabla_b \nabla_a u^a \nonumber \\
&=& \nabla_a(u^b \nabla_b u^a) - (\nabla_au^b)(\nabla_b u^a)
   -\nabla_b(u^b \nabla_a u^a) + (\nabla_b u^b)^2 \nonumber \\
&=& \nabla_i(Ku^i +a^i) - K_{ab}K^{ab} + K_a^a K^b_b
\label{rabcd}
\end{eqnarray}
(Note that for $K_{ij} = K_{ji} = -\nabla_i u_j - u_i a_j$,
we have $K \equiv K^i_i = -\nabla_i u^i$ and
$K_{ij} K^{ij} = (\nabla_iu^j) (\nabla_ju^i)$).
Further using
\begin{equation}
R = -R~g_{ab} u^a u^b = 2 (G_{ab}-R_{ab}) u^a u^b ~,
\label{rinhone}
\end{equation}
and the identity
\begin{equation}
2~G_{ab} u^a u^b = K_a^a K^b_b   - K_{ab}K^{ab} + {}^{(3)}\mathcal{R}~,
\label{ginr}
\end{equation}
we can write the scalar curvature as
\begin{equation}
R = {}^{(3)}\mathcal{R} +K_{ab}K^{ab} - K_a^a K^b_b - 2 \nabla_i (Ku^i + a^i)
  \equiv L_{\rm ADM} -2 \nabla_i (Ku^i + a^i) ~,
\label{rinh}
\end{equation}
where $L_{\rm ADM}$ is the ADM Lagrangian.
This is the result used in the article.
   
Let us now integrate (\ref{rinh}) over a four volume $\mathcal{V}$
bounded by two space-like hypersurfaces $\Sigma_1$ and $\Sigma_2$
and a time-like hypersurface $\mathcal{S}$.
The space-like hypersurfaces are constant time slices with normals $u^i$,
and the time-like hypersurface has normal $n^i$ orthogonal to $u^i$.
The induced metric on the space-like hypersurface $\Sigma$ is
$h_{ab} = g_{ab} + u_a u_b$, while the induced metric on the time-like
hypersurface $\mathcal{S}$ is $\gamma_{ab} = g_{ab} - n_a n_b$.
$\Sigma$ and $\mathcal{S}$ intersect along a 2-dimensional surface
$\mathcal{Q}$, with the induced metric
$\sigma_{ab} = h_{ab} - n_a n_b = g_{ab} + u_a u_b - n_a n_b$. 
With $g_{00}=-N^2$, we get
\begin{eqnarray}
A_{\rm EH} &=& {1 \over 16\pi} \int_\mathcal{V} d^4x~\sqrt{-g}~R \nonumber \\
  &=& {1 \over 16\pi} \int_\mathcal{V} d^4x~\sqrt{-g}~L_{\rm ADM}
    -{1 \over 8\pi} \int_{\Sigma_1}^{\Sigma_2} d^3x~\sqrt{h}~K
    - {1 \over 8\pi} \int_\mathcal{S}
      dt~d^2x~N~\sqrt{\sigma}(n_ia^i) ~.
\label{bigeqn}
\end{eqnarray}
 
Let the hypersurfaces $\Sigma, \mathcal{S}$ as well as their intersection
2-surface $\mathcal{Q}$ have the corresponding extrinsic curvatures
$K_{ab}, \Theta_{ab}$ and $q_{ab}$. 
In the literature, the Einstein-Hilbert action is conventionally expressed
as a term having only the first derivatives, plus an integral of the trace
of the extrinsic curvature over the bounding surfaces.
It is easy to obtain this form using the foliation condition $n_i u^i=0$
between the surfaces, and noting
\begin{equation}
n_i a^i = n_i u^j \nabla_j u^i = -u^j u^i \nabla_j n_i
= (g^{ij} - h^{ij}) \nabla_jn_i =- \Theta + q ~,
\label{thetaink}
\end{equation} 
where $\Theta\equiv\Theta^a_a$ and $q\equiv q^a_a$ are the traces of the
extrinsic curvature of the surfaces, when treated as embedded in the
4-dimensional or 3-dimensional enveloping manifolds.
Using (\ref{thetaink}) to replace $(n_i a^i)$ in the last term of
(\ref{bigeqn}), we get the result   
\begin{eqnarray}
A_{\rm EH} &+& {1 \over 8\pi} \int_{\Sigma_1}^{\Sigma_2} d^3x~\sqrt{h}~K
 - {1 \over 8\pi} \int_\mathcal{S} dt~d^2x~N~\sqrt{\sigma}~\Theta \nonumber \\
&=& {1 \over 16\pi} \int_\mathcal{V} d^4x~\sqrt{-g}~L_{\rm ADM}
 - {1 \over 8\pi} \int_\mathcal{S} dt~d^2x~N~\sqrt{\sigma}~q ~.
\label{EHtoADM}
\end{eqnarray}
In the first term on the right hand side,
$L_{\rm ADM}$ contains ${}^{(3)}\mathcal{R}$,
which in turn contains second derivatives of the metric tensor.
The second term on the right hand side removes these second derivatives
making the right hand side equal to the $\Gamma^2$-action for gravity.
On the left hand side, the second and third terms are integrals of the
extrinsic curvatures over the boundary surfaces, which when added to the
Einstein-Hilbert action give the quadratic action without second derivatives.
This is the standard result often used in the literature,
which---unfortunately---misses the importance of the $(n_i a^i)$ term
in the action by splitting it.

\section{Possible Boundary Terms in the Effective Action for Gravity
\label{boundaryterms}}

For any field theory, its symmetry principles constrain the structure
of terms that may appear in its action.
In case of gravity, the symmetry principle is general covariance.
When the space-time has no boundary, allowed terms in the Lagrangian
density for gravity are invariant scalars formed from the metric
$g_{ij}$ and the covariant derivative operator $\nabla_k$.
The action is obtained by integrating these invariant scalars over the
the invariant space-time measure $d^4x \sqrt{-g}$.
When the effective action is restricted to contain no more than two
derivative operators, there are only two possible terms,
a constant and $R$, and the action takes the form
\begin{equation}
A_{\rm bulk} \equiv  \int d^4x~\sqrt{-g}~(c_1+c_2R) ~=
{1\over 16\pi G} \int d^4x~\sqrt{-g}~(R-2\Lambda) ~,
\end{equation}  
which is the Einstein-Hilbert action with a cosmological constant.
(No bulk term involving the totally antisymmetric tensor
$\epsilon_{ijkl}$ is possible, when there is no torsion.)
The constants $c_1$ and $c_2$ are traded of for $G$ and $\Lambda$,
but their values---even their signs---cannot be ascertained
without further physical inputs. 

When the space-time has a boundary,
additional variables describing the geometry of the boundary
can also appear in the action, giving rise to new terms.
The form of these new terms can still be restricted to a great
extent by general considerations as described below.

\subsection{Euclidean gravity}

The structure of the boundary terms is easier to understand
in Euclidean space-time. Let $\mathcal{M}$ be a compact region
of Euclidean space-time with the boundary $\partial\mathcal{M}$.
The geometry of a closed orientable boundary can be fully described
by its outward pointing unit normal, $w^a$, with $w^aw_a=1$.
Generically the boundary is specified in a coordinate dependent way
(e.g. by setting some coordinate to a constant value).
Then general coordinate transformations move the boundary around,
and $w^a$ is not a generally covariant vector,
although it is Lorentz covariant (with appropriate Euclidean meaning).
In such a situation, possible terms in the effective action for gravity
can be obtained by fictitiously treating $w^a$ as a generally covariant
vector, and constructing all possible invariant scalars.
This prescription ensures that symmetries of the effective action
are violated only through $w^a$, and not through any other variable.
(For example, a similar prescription is used to construct effective
chiral Lagrangians of strong interactions. The chirally non-invariant
mass parameter is assigned a fictitious transformation property that
cancels the chiral transformation of hadron fields,
and then all possible chirally invariant terms are written down.)

Since $w^a$ is defined only on the boundary,
the action terms containing it have to be boundary integrals.
Since we would like to have a local Lagrangian description of the theory,
a foliation may be used to extend the value of $w^a$ through the whole
space-time. Such a foliation is by no means unique, and the action
terms must not depend on how the extension of $w^a$ is carried out.
This requirement is guaranteed by demanding that action terms
containing $w^a$ be total divergences; Gauss's law then implies that
$w^a$ defined on the boundary is sufficient to evaluate these terms.
Thus the boundary terms in the effective action for gravity take the form
\begin{equation}
A_{\rm boundary} = ({\rm constant}) \int d^4x \sqrt{g}~\nabla_a V^a,
\end{equation}
where $V^a$ is a vector constructed from $g_{ij}$, $\nabla_k$ and $w_b$.
This total divergence term does not affect the dynamics of classical
gravity, but it may affect the dynamics of the quantum theory.

With the restriction of no more than two derivative operators in the action, 
the possible candidates for $V^a$ are:
(i) with zero derivatives, $V^a=w^a$;
(ii) with one derivative, 
$V^a=(w^a \nabla_b w^b, w^b \nabla_b w^a, w_b \nabla^a w^b)$.
Of these, $w_b \nabla^a w^b = 0$, due to the normalization $w^b w_b=1$.
Furthermore, the term $\nabla_j (w^b \nabla_b w^j)$ involving the
``acceleration" $a^j= w^b \nabla_b w^j$ integrates to zero,
since use of Gauss's law converts the integrand to $w_ja^j$
which vanishes identically.
(One may also consider the possibility,
$V^a = (g)^{-1/2} \epsilon^{abcd} w_b \nabla_c w_d$,
but it integrates to zero up on using Gauss's law.)
Thus we are left with just two possibilities $V^a=(w^a,w^a \nabla_b w^b)$.
In terms of the induced metric on $\partial\mathcal{M}$, $f_{ab}$,
the corresponding contributions to the action are:
\begin{equation}
A_{\rm boundary}^{(1)}
= \int_{\mathcal{M}} d^4x \sqrt{g}~\nabla_a w^a
= \int_{\partial\mathcal{M}} d^3x \sqrt{f}
= {\rm Vol}(\partial\mathcal{M}) ~,
\label{volumeterm}
\end{equation}
\begin{equation}
A_{\rm boundary}^{(2)}
= \int_{\mathcal{M}} d^4x \sqrt{g}~\nabla_a (w^a \nabla_b w^b)
= \int_{\partial\mathcal{M}} d^3x \sqrt{f}~\nabla_b w^b ~
= -\int_{\partial\mathcal{M}}d^3x \sqrt{f}~K_{\rm ext} ~.
\end{equation}
The former is the total volume of the boundary,
while the latter is the integral of the extrinsic curvature
$K_{\rm ext}=-\nabla_b w^b$ over the boundary.
The most general lowest order effective action
for Euclidean gravity thus takes the form
\begin{equation}
A_{\rm grav} \equiv \int_{\mathcal{M}} d^4x~\sqrt{g}~(c_1+c_2R)
  + \int_{\partial\mathcal{M}} d^3x \sqrt{f}~(c_3+c_4 K_{\rm ext}) ~.
\label{euclsurface}
\end{equation}  
The constants appearing here have the dimensions
$c_1\sim L^{-4}, c_2\sim L^{-2}, c_3\sim L^{-3}, c_4\sim L^{-2}$.

For a closed but only piecewise smooth boundary,
it is convenient to separate $A_{\rm boundary}$ in to the contribution
from the smooth part and the contribution from the edges.
The normal to the boundary is discontinuous in going across the edge,
and so the gradient of the normal is singular along the edge.
This singularity does not contribute to $A_{\rm boundary}^{(1)}$,
but it does contribute to $A_{\rm boundary}^{(2)}$.
The edge contribution to $A_{\rm boundary}^{(2)}$ can be evaluated
by rounding off the edge in a limiting procedure.
In this limit, the curvature cancels with the corresponding factor from
the integration measure, and only the angular discontinuity $\delta$
of the normal across the edge is left behind.
Let $\mathcal{Q}$ be the set of edges with the induced metric $\sigma_{ab}$.
Then
\begin{equation}
-\int_{\partial\mathcal{M}}d^3x \sqrt{f}~K_{\rm ext} ~\longrightarrow~
-\int_{\partial\mathcal{M},{\rm smooth}}d^3x \sqrt{f}~K_{\rm ext}
+\int_{\mathcal{Q}}d^2x \sqrt{\sigma}~\delta ~.
\end{equation}
A common situation is the one where the unit normals to the boundary
on either side of the edge, $w^{(1)a}$ and $w^{(2)a}$, are orthogonal.
The edge contribution involves both, in the form of a double divergence.
The first divergence embeds the 3-dimensional boundary in the 4-dimensional
space-time and then the second divergence embeds the 2-dimensional edge
in the 3-dimensional boundary.
With $D$ denoting the projection of $\nabla$ on to the boundary
$\partial\mathcal{M}$, the edge contribution becomes
\begin{eqnarray}
A_{\rm boundary}^{(2),{\rm edge}}
= {\pi\over2}
  \int_{\mathcal{M}} d^4x \sqrt{g}~\nabla_a (w^{(1)a} f_{bc} D^b w^{(2)c})
= {\pi\over2} \int_{\partial\mathcal{M}} d^3x \sqrt{f}~f_{bc} D^b w^{(2)c}
= {\pi\over2} \int_{\mathcal{Q}} d^2x \sqrt{\sigma} ~,
\end{eqnarray}
which is the total area of the edge surface multiplied by the angular
discontinuity $\delta=\pi/2$.

In dealing with horizons in Euclidean space-time,
a particularly useful geometry is that of a punctured manifold.
This geometry can be obtained by taking the limit where
the boundary reduces to a point in a two-dimensional submanifold,
while the other transverse dimensions remain unaffected.
The boundary action is then proportional to the horizon area:
\begin{equation}
\label{puncturedmanifold}
\int_{\partial\mathcal{M}}d^3x \sqrt{f}~(c_3+c_4 K_{\rm ext})
~\longrightarrow~ c_4\theta\int_{\mathcal{Q}}d^2x \sqrt{\sigma} ~,
\end{equation}
where $\theta$ is the conical angle corresponding to the puncture.
For a puncture without conical singularity, $\theta=2\pi$,
consistent with Euler's theorem.

If the boundary can be analytically continued back to Minkowski
space-time in some suitable coordinates, then we will have
analogous boundary terms in Minkowski space-time also.
If we impose the condition that the complete action does not contain
derivatives higher than the first, then we must have $c_4=2c_2$.
We see from equation (\ref{EHtoADM}) that in this case the sum of
bulk and surface terms lead to the quadratic $\Gamma^2$-action.
We stress that the ADM action contains second derivatives of $g_{ab}$
with respect to the spatial coordinates (through ${}^{(3)}R$),
while the $\Gamma^2$-action has no second derivatives at all.
In Euclidean space-time, both space and time are treated on an equal
footing, and any covariant prescription which removes the second
derivatives along one direction will remove them along all directions,
leading to the $\Gamma^2$-action.

\subsection{Minkowski space-time with foliation}

It is possible to figure out the possible boundary terms
even for a space-time with Minkowski signature,
in presence of pre-specified foliations.
In this case, the boundaries typically have
both space-like ($\Sigma$) and time-like ($\mathcal{S}$) parts
as described in Appendix \ref{actionstructure}.
The corresponding normals $u^a$ and $n^a$ satisfy $u_au^a=-1, n_an^a=1$,
and $u^an_a=0$ on the intersection of $\Sigma$ and $\mathcal{S}$.
$\Sigma$ and $\mathcal{S}$ have induced metrics $h_{ab}=g_{ab}+u_a u_b$
and $\gamma_{ab}=g_{ab}-n_a n_b$, and extrinsic curvatures
$K=-\nabla_a u^a$ and $\Theta=-\nabla_a n^a$ respectively.
Now we need to make a clear distinction between two different situations: 

(a) We may insist that $u^a$ is given only on $\Sigma$, and $n^a$ is
given only on $\mathcal{S}$ with arbitrary extensions elsewhere.
Then the surface terms we obtain should not depend
on the manner in which these are extended.
This situation is similar to that of the Euclidean case discussed above,
with one crucial difference:
The surface of intersection of $\Sigma$ and $\mathcal{S}$
(i.e. $\mathcal{Q}$) cannot be smoothly rounded off because the two
normals $u^a$ and $n^a$ have normalizations of opposite signs.
With unrelated coefficients for the space-like and the time-like parts of
the boundary, the boundary action becomes
\begin{equation}
A_{\rm boundary}
= k_1{\rm Vol}(\Sigma)
+ k_2\int_\Sigma d^3x \sqrt{h}~K
+ k_3{\rm Vol}(\mathcal{S})
+ k_4\int_\mathcal{S}dtd^2x \sqrt{|\gamma|}~\Theta
+ k_5{\rm Area}(\mathcal{Q}) ~.
\label{finalnofoli}
\end{equation}
If a smooth analytic continuation between Minkowski and Euclidean
space-times is assumed, then the coefficients are related according to:
$k_3=-k_1, k_4=-k_2, k_5=(\pi/2)k_2$.

(b) It is, however, commonplace to assume a more elaborate geometry, viz.
that there exists a \emph{foliation} of space-time by space-like surfaces, 
for which $u^a(x)$ is the normal.
For example, the ADM formulation explicitly uses such a foliation.
In such a case, the possible terms in the action can depend on the
complete vector field $u^a(x)$ and not just its value on the boundary.
The resultant action would have more parameters, and hence would be less
predictive.
In general, the space-time structure need not admit a foliation
in terms of time-like surfaces, and we will continue to assume
that $n^a$ is specified \emph{only} on $\mathcal{S}$.
Our final results should then depend only on the value of $n^a$ on the
boundary, and not in the manner in which it may be extended elsewhere.

Let us proceed as in the case of Euclidean space-time,
pretending that $u^a$ and $n^a$ can be extended
in some sensible fashion to the bulk as two vector fields.
With $u^a(x)$ and $n^a(x)$ treated as genuine vector fields,
one can write down several new terms for the bulk action.
We, however, focus on total divergence boundary terms only.
These terms are again integrals of the form $\nabla_a V^a$,
where $V^a$ is built from $g_{ij}, \nabla_k, u^b, n^c$,
with the derivative acting at most once.  
The following vectors exhaust the possibilities:
\begin{eqnarray}
V^a = &(& u^a, u^a\nabla_b u^b, u^b\nabla_b u^a,
               u^a\nabla_b n^b, u^b\nabla_b n^a, \nonumber \\
       && n^a, n^a\nabla_b n^b, n^b\nabla_b n^a,
               n^a\nabla_b u^b, n^b\nabla_b u^a, \nonumber \\
       && u^b\nabla^a n_b, u^au^bu^c\nabla_b n_c, u^an^bu^c\nabla_b n_c,
                           n^au^bu^c\nabla_b n_c, n^an^bu^c\nabla_b n_c\ ) ~.
\label{listofv}
\end{eqnarray}
The second line is obtained from the first, by the obvious interchange
$u^a \longleftrightarrow n^a$. 
For the third line, there is no need to include  such interchanges,
 because $u^b\nabla^an_b = -n_b\nabla^au^b$.
The first three terms of both first and second lines have already been
discussed in the previous subsection, and the remaining terms arise
because of the existence of the second vector field.
When $\nabla_a V^a$ is integrated over a space-time region,
we get (i) boundary integral of $u_a V^a$ over $\Sigma$,
and (ii) boundary integral of $n_a V^a$ over $\mathcal{S}$. 

These dot products are given by
\begin{equation}
u_a V^a = (1, \nabla_au^a, 0, \nabla_an^a, u^au^b\nabla_bn_a,\ \ 
           0, 0, u^an^b\nabla_bn_a, 0, 0,\ \ 
           u^au^b\nabla_an_b, u^au^b\nabla_an_b, n^au^b\nabla_an_b, 0, 0) ~,
\label{udotv}         
\end{equation}
and
\begin{equation}
n_a V^a = (0, 0, n^au^b\nabla_bu_a, 0, 0,\ \ 
           -1, -\nabla_an^a, 0, -\nabla_au^a, n^an^b\nabla_bu_a,\ \ 
           n^au^b\nabla_an_b, 0, 0, -u^au^b\nabla_an_b, -n^au^b\nabla_an_b) ~.
\label{ndotv}         
\end{equation}
Together, we have to consider the set of scalars
\begin{eqnarray}
(u_aV^a, n_aV^a) =
(1, \nabla_au^a, \nabla_an^a, u^an^b\nabla_bn_a, n^au^b\nabla_bu_a) ~,
\label{fullset}
\end{eqnarray}
where we have used the condition $n_au^a=0$ to eliminate some of the
scalars in favour of the ones listed.

Let us begin with possible integrals over the boundaries $\Sigma$.
We first note that the vector $n^a$ is \emph{not} specified over most
of the surface (it exists only because we arbitrarily extended it
for convenience), but it is known on the 2-surface $\mathcal{Q}$.
This, in turn, implies that we cannot have the two terms
$(u^an^b\nabla_bn_a,  n^au^b\nabla_bu_a)$ integrated over $\Sigma$,
since their values depend on how $n^a$ is extended.
The $\nabla_a n^a$ term requires more care, however.
Note that the projection $D_an^a$ of $\nabla_an^a$ on to $\Sigma$
is given by
\begin{equation}
D_an^a \equiv (\delta^i_a+u^iu_a) (\delta_j^a+u_ju^a) \nabla_in^j
       = \nabla_an^a +u^iu^j\nabla_in_j ~.
\label{dproject}
\end{equation}
Therefore, $\nabla_an^a$ differs from $h^{ab}D_an_b$
only by the term $u^au^b\nabla_an_b = -n_bu^a\nabla_au_b$,
which has already been considered.
The integral of $h^{ab}D_an_b$ over $\Sigma$ is allowed,
because Gauss's law converts it in to an integral over $\mathcal{Q}$
on which $n^a$ is known.
(In other words, the integral of either of the two terms on the right
hand side of (\ref{dproject}) depends on the manner in which $n_a(x)$
is extended in to $\Sigma$, but their combined integral depends only
on the known value of $n_i$ on $\mathcal{Q}$.)
The corresponding boundary action is
\begin{equation}
A_{\rm boundary}^{(3)}
= -\int_\Sigma d^3x \sqrt{h}~h_{bc} D^b n^c
= -\int_{\mathcal{Q}} d^2x \sqrt{\sigma},
\end{equation}
which is the total area of the surface $\mathcal{Q}$.
Thus the total contribution of terms in (\ref{fullset}) on the boundary
$\Sigma$ is
\begin{equation}
A_{\rm boundary:\Sigma}
= \int_\Sigma d^3x \sqrt{h}~(k_1+k_2 K) 
  + k_5\int_{\mathcal{Q}} d^2x \sqrt{\sigma}
= k_1{\rm Vol}(\Sigma) + k_2\int_\Sigma d^3x \sqrt{h}~K
  + k_5{\rm Area}(\mathcal{Q})
\end{equation}
If the total action has to contain only up to first {\it time} derivatives,
then $k_2=2c_2$. 

Let us next consider the contribution of the terms in (\ref{fullset})
on the boundary $\mathcal{S}$.
The first three terms can be analyzed just as in case of $\Sigma$, yielding
\begin{equation}
A_{\rm boundary:\mathcal{S}}
= \int_{\mathcal{S}}  d^3x \sqrt{|\gamma|}~(k_3+k_4 \Theta) 
  + k_5\int_{\mathcal{Q}} d^2x \sqrt{\sigma}
= k_3{\rm Vol}(\mathcal{S}) + k_4\int_{\mathcal{S}}  d^3x \sqrt{|\gamma|}~\Theta
  + k_5{\rm Area}(\mathcal{Q})
\end{equation}
In addition, since $u^a(x)$ is a vector field defined everywhere in
$\mathcal{S}$, we can no longer ignore the last two terms
$(u^an^b\nabla_bn_a, n^au^b\nabla_bu_a)$
while integrating over $\mathcal{S}$. (These are the
``accelerations" of one normal dotted with the other normal.)
Of these, the term $n^au^b\nabla_bu_a=n^ia_i$ leads to the integral of the
surface gravity, and has been extensively discussed earlier in this paper:
\begin{equation}
A_{\rm boundary}^{(4)}
= k_6\int_\mathcal{S}dtd^2x \sqrt{|\gamma|}~n^ia_i
= k_6\int_\mathcal{S}dtd^2x N\sqrt{\sigma}~n^ia_i ~.
\label{ndota}
\end{equation}
What remains is the term
\begin{equation}
A_{\rm boundary}^{(5)}
= k_7\int_\mathcal{S}dtd^2x \sqrt{|\gamma|}~u^an^b\nabla_bn_a
=-k_7\int_\mathcal{S}dtd^2x N\sqrt{\sigma}~n^an^b\nabla_bu_a ~.
\label{mystery}
\end{equation}
To understand the nature of this term, consider a coordinate system
in which $x^1=$ constant corresponds to the surface $\mathcal{S}$,
and the metric has the form $g_{00}=-N^2, g_{11}=M^2, g_{0\alpha}=0,
g_{1A}=0, g_{AB}=\sigma_{AB}$ with $A,B=2,3$.
Then, a simple calculation shows that
$n^an^b\nabla_bu_a=(MN)^{-1}(\partial M/\partial t)$.
This term vanishes if $(\partial M/\partial t)=0$.
In general, its contribution is
\begin{equation}
A_{\rm boundary}^{(5)}
= -k_7\int_\mathcal{S}dtd^2x N\sqrt{\sigma}~\frac{\dot M}{MN}
= -k_7\int_\mathcal{Q}d^2x \sqrt{\sigma}~(\ln M) \bigg|_{t_1}^{t_2} ~.
\end{equation}

To summarize, equation (\ref{finalnofoli}) gives the total boundary action,
when $u^i$ and $n^i$ are defined only on the boundary and no foliation is
assumed.
This result agrees with the one obtained in case of Euclidean space-time.
If the vector field $u^a(x)$ is related to a space-time foliation, then
two additional terms appear, given by (\ref{ndota}) and (\ref{mystery}).
This concludes our general analysis of boundary terms.

\subsection{Boundary terms for a horizon}

When the boundary is a horizon, and not just any hypersurface,
the possible terms in the effective action get restricted further.
This happens because the integration measure in time direction is
$Ndt$, and with $N=0$ on the horizon some of the integrals vanish.
Among the various contributions in (\ref{finalnofoli}),
(i) The term ${\rm Vol}(\mathcal{S})$ vanishes.
(ii) In the term with $\Theta$, we can use (\ref{thetaink}) to write
$\Theta=q-n^ia_i$, and note that the integral of $q$ vanishes.
The $n^ia_i$ term, which of course is the same as in (\ref{ndota}),
leads to the area of the horizon when $N\to0$.
(iii) Finally, the term in (\ref{mystery}) vanishes when
$N\to0, MN=$constant, or when $\dot{M}=0$ on the horizon.
Thus, when the boundary is a horizon,
the most general boundary action is of the form
\begin{equation}
A_{\rm boundary: horizon}
= k_1{\rm Vol}(\Sigma) + k_2\int_\Sigma d^3x \sqrt{h}~K
+ k_4{\rm Area}(\mathcal{Q}) ~.
\end{equation}
The result obtained with the ADM action in the paper corresponds to
$k_1=0, k_2=2c_2,k_4=4\pi c_2$.
In the special case of a static horizon,
the space-like surfaces $\Sigma_1$ and $\Sigma_2$ are eliminated
by the periodic boundary conditions in time direction, leaving behind
\begin{equation}
A_{\rm boundary: static~horizon} = k_4{\rm Area}(\mathcal{Q}) ~.
\end{equation}


\begin{thebibliography}{99}
\bibitem{comment0} It is assumed that each curve in the congruence is
         extended to the maximum possible value of the affine parameter,
         say, the proper time of clocks moving along the curves.
         If the space-time has real physical singularities
         and the curves in the congruence hit the singularity,
         then one cannot extend the curve further.
         But that is irrelevant for our purpose, and the range
         we are interested in is usually obvious from the context.
\bibitem{comment1} Integration over null surfaces like horizons requires
         some care, and can be handled using a limiting procedure.
         One also needs to be careful in handling the intersection
         of future and past space-like surfaces $\Sigma_2$ and $\Sigma_1$
         on the bifurcation sphere.
         Though we have not discussed these issues in detail, they can be
         handled rigorously and do not affect our arguments or results.
\end{thebibliography}
\end{document}